\documentclass[prd,twocolumn,superscriptaddress,nofootinbib]{revtex4-2}
\usepackage{amsmath,amssymb,graphicx}
\usepackage{url}
\newcommand{\E}{\mathbb{E}}
\newcommand{\K}{\mathcal{K}}
\newcommand{\vz}{\mathbf z}
\newcommand{\vy}{\mathbf y}
\newcommand{\vk}{\mathbf k}
\newcommand{\vrr}{\mathbf r}
\newcommand{\dd}{\mathrm d}
\newcommand{\Sd}{\dot S}
\newcommand{\I}{\mathcal{I}}
\renewcommand{\Re}{\operatorname{Re}}
\begin{document}
\title{A Poisson-gas approach to medium-induced gluon radiation}
\author{Manoel Rodr\'iguez Calvo}
\affiliation{Escuela Internacional de Doctorado (EIDUNED), Universidad Nacional de Educaci\'on a Distancia (UNED), 28040 Madrid, Spain}
\begin{abstract}
The precision era of jet-quenching phenomenology requires a quantitative control of medium-induced gluon radiation.
We consider the BDMPS-Z spectrum in the soft-gluon limit for a Yukawa (Gyulassy--Wang) potential.
Writing the Yukawa dipole cross section as a Laplace superposition of Gaussians, we express the fully resummed spectrum as an
average over Poisson-distributed configurations of Gaussian scattering centers, each of which is solved in closed form.
This yields a master formula for the spectrum both differential in the transverse momentum $k_\perp$ and integrated with a cut
on $k_\perp$, valid for static and expanding media with arbitrary profiles. Its evaluation by Monte Carlo sampling of the
configurations is efficient and trivially parallelizable. The method reproduces the first order in opacity and published
results without kinematic cuts, while with a cut we find sizeable relative deviations from previous results. We argue that the same construction extends to the more complex building blocks of an in-medium parton cascade,
such as finite energy fraction and emissions off color antennas, and provides a natural starting point for analytical
approximations. The code is publicly available.
\end{abstract}\maketitle

\section{Introduction}\label{sec:intro}

The analysis of hard probes in high-energy heavy-ion collisions is one of the main experimental handles on the properties of the quark--gluon plasma created in those collisions. One of the central theoretical tools to describe hard probes in such an environment is the medium-induced gluon radiation spectrum. With more precise measurements and new observables becoming available, a precise evaluation of this spectrum has become essential.

In this paper we consider the BDMPS-Z formalism \cite{BDMPS1997a,BDMPS1997b,Zakharov1996,Zakharov1997} for gluon radiation off an eikonal parton in the soft limit. In this description the medium acts as a color background field with which the jet interacts. We neglect recoil effects and assume the background field to be classical and to obey Gaussian statistics. The medium is then characterized by the two-point function
\begin{equation}
\langle A^{-}_{a}(x^+,\mathbf x)\,A^{-}_{b}(y^+,\mathbf y)\rangle=\delta_{ab}\,n(x^+)\,\delta(x^+-y^+)\,\gamma(\mathbf x-\mathbf y),
\end{equation}
where $\mathbf x$ and $\mathbf y$ are transverse positions, $n(s)$ is the density of scattering centers at path length $s$, and $\gamma$ is the Fourier transform of the elastic scattering potential, whose screening mass $\mu(s)$ may also depend on $s$. In the soft limit the medium enters through an adjoint dipole, whose Gaussian average is $e^{-\frac12\int\dd s\,n(s)\sigma(\mathbf r)}$ with $\sigma(\mathbf r)=2g^2N_c[\gamma(\mathbf 0)-\gamma(\mathbf r)]=\int\frac{\dd^2q}{(2\pi)^2}V(q)\,(1-e^{i\mathbf q\cdot\mathbf r})$; that is, $V(q)=2g^2N_c\tilde\gamma(q)$, with $\tilde\gamma$ the Fourier transform of $\gamma$, is the elastic collision rate of Ref.~\cite{AndresApolinarioDominguez2020}.

Typically, the potential is taken to be of Yukawa type (Gyulassy--Wang model) \cite{GyulassyWang1994}. Another option is the HTL potential \cite{AurencheGelisZaraket2002}, which shares the $1/q^4$ tail of the Yukawa potential and, when regularized with a magnetic mass, also admits the representation as a Poisson gas of Gaussian scattering centers used in this paper (Appendix~\ref{app:htl}). In this paper we focus on the Yukawa potential and leave other cases for future studies. This tail is associated with rare but relevant hard kicks. It makes the dipole cross section behave as $r^2\log(1/\mu^2r^2)$ at small $r$, and it has proven a hurdle that is difficult to overcome analytically. Traditionally, two analytical limits have been used: GLV \cite{GLV2001,GLV2000,Wiedemann2000} and the harmonic oscillator (HO) \cite{BDMPS1997a,SalgadoWiedemann2003}. The GLV spectrum is the first order of an expansion in the opacity of the medium. The HO approximation keeps only the quadratic term of the dipole cross section, which amounts to a fixed width for every scattering, but it resums multiple scatterings. More recently, the improved opacity expansion (IOE) \cite{MehtarTani2019,MehtarTaniTywoniuk2020,BarataMehtarTani2020}, which expands around the HO, and the resummed opacity expansion (ROE) \cite{IsaksenTakacsTywoniuk2023}, which extends the description to the Bethe--Heitler regime, have made significant progress.

From the numerical point of view, Andr\'es \textit{et al.} \cite{AndresApolinarioDominguez2020} proposed to compute the spectrum by solving a PDE, following earlier work by Caron-Huot and Gale \cite{CaronHuotGale2010}, and later extended this approach to expanding media with kinematic cuts \cite{Andres2024}. Closer to the approach followed in this paper, Feal and V\'azquez \cite{FealVazquez2018} used a Monte Carlo method. Both approaches resum multiple scatterings to all orders.

In this paper we also tackle the problem in full generality. We write the Gyulassy--Wang dipole cross section as a mixture of Gaussians, which allows us to interpret the medium as a Poisson gas of Gaussian scattering centers, each characterized by its position and its transverse-momentum width (see Sec.~\ref{sec:representation}). Analytical calculations lead us to a master formula that encompasses both the spectrum differential in $k_\perp$ and the spectrum with a cut on $k_\perp$. This master formula is amenable to a standard Monte Carlo integration over configurations, which is efficient, highly parallelizable and able to handle arbitrary profiles of both $\mu(s)$ and $n(s)$.

Unlike PDE-based approaches, each configuration of the medium is solved in closed form. This provides a clearer picture of the BDMPS-Z spectrum: every configuration is Gaussian and is solved like the harmonic oscillator, with a frequency given by a sum of impulses located at the scattering centers. Feal and V\'azquez also use a Monte Carlo method, but they sample momentum kicks in discretized time, whereas we sample Poisson-distributed centers in continuous time. On the other hand, Feal and V\'azquez include exact kinematics, large emission angles, a thermal gluon mass and the external legs. We leave the inclusion of this physics in our formalism for future work.

The method reproduces the first order in opacity and the published results without kinematic cuts. With a cut, we find sizeable relative deviations from previous results, both at small gluon energy fraction, where the cut strongly suppresses the spectrum, and at large energy fraction, where they reach 5--20\,\%. The code is publicly available at \url{https://github.com/manoel-maker/poisson-gas-bdmps_z}.

The paper is organized as follows. In Sec.~\ref{sec:representation} we set up the calculation: starting from known expressions for the BDMPS-Z spectrum \cite{Wiedemann2000,AndresApolinarioDominguez2020} and relying on our representation of the medium, we write the spectrum in a form from which the master formula, on which this paper is based, follows directly. In Sec.~\ref{sec:master} we derive the master formula for the spectrum differential in $k_\perp$ and with a cut, for arbitrary $\mu(s)$ and $n(s)$, by integrating the transverse positions center by center; the equivalent route through Arnold's propagator~\cite{Arnold2009} is given in Appendix~\ref{app:arnold}. In Sec.~\ref{sec:integrated} we obtain the spectrum integrated over all $k_\perp$, again by two routes. In Sec.~\ref{sec:mc} we present the Monte Carlo method. Although the methodology is standard, we give a self-contained review: we present Mecke's formula and discuss the importance sampling of the time between the emission and a marked scattering center. Section~\ref{sec:validation} contains the validation and the comparison with published results, and Sec.~\ref{sec:summary} our summary and outlook. Appendix~\ref{app:htl} discusses the HTL potential with a magnetic mass.

\section{Representation of the medium}\label{sec:representation}

\subsection{Starting point}\label{sec:start}
As detailed in the Introduction, we study the medium-induced gluon emission spectrum in the soft-gluon limit. Our starting point is the
spectrum differential in $\vk$, Eqs.~(2.2)--(2.4) of Ref.~\cite{AndresApolinarioDominguez2020}, written in position space:
\begin{multline}\label{eq:start}
\omega\frac{\dd I}{\dd\omega\,\dd^2k}=\frac{2\alpha_sC_R}{(2\pi)^2\omega^2}\,\Re\int_0^\infty\!\dd t_2\int_0^{t_2}\!\dd t_1\\
\times\int\!\dd^2z\;e^{-i\vk\cdot\vz}\,e^{-\frac12\int_{t_2}^\infty \dd s\, n(s)\sigma(z)}\\
\times\nabla_{\vy}\!\cdot\nabla_{\vz}\,\K(t_2,\vz;t_1,\vy)\Big|_{\vy=0}
\;-\;(\text{vacuum}),
\end{multline}
with
\begin{multline}\label{eq:K}
\K(t_2,\vz;t_1,\vy)=\int_{\vrr(t_1)=\vy}^{\vrr(t_2)=\vz}\mathcal D\vrr\\
\times\exp\Big\{\int_{t_1}^{t_2}\!\dd s\Big[\frac{i\omega}{2}\dot\vrr^{\,2}-v(\vrr(s),s)\Big]\Big\},
\end{multline}
with $v=\tfrac12 n\sigma$, and ``(vacuum)'' denotes the same expression with $n\equiv0$. In Eq.~\eqref{eq:start} the medium enters through two factors. The first
one is the kernel $\K$, which describes the evolution of the dipole between $t_1$ and $t_2$. The second factor is the broadening factor
$e^{-\frac12\int_{t_2}^\infty n\sigma(z)}$. It has the form of a kernel without kinetic term: after $t_2$ the kinetic phases of the
amplitude and of the conjugate amplitude cancel, the dipole size is frozen at $\vz$, and the factor describes the broadening of the
emitted gluon. That the problem reduces to a single dipole is a consequence of the soft limit, in which the emitter is purely eikonal.

For the Yukawa potential $V(q)=8\pi\mu^2/(q^2+\mu^2)^2$ \cite{AndresApolinarioDominguez2020}, with a local screening mass $\mu(s)$ \cite{Andres2024}, the dipole cross section
$\sigma(r)=\int\frac{\dd^2q}{(2\pi)^2}\,V(q)\,(1-e^{i\mathbf q\cdot\vrr})$ gives
\begin{equation}\label{eq:dipole}
v(r,s)=\tfrac12 n(s)\sigma(r)=n(s)\,\big[1-\mu(s) r\,K_1(\mu(s) r)\big].
\end{equation}

\subsection{The Yukawa dipole as a superposition of Gaussians}\label{sec:laplace}
Since $V(q)$ is a rational function of $q^2+\mu^2$, its inverse Laplace transform~\cite{Widder1941} is immediate,
\begin{equation}\label{eq:schwinger}
\frac{1}{(q^2+\mu^2)^2}=\int_0^\infty\!\dd\lambda\;\lambda\,e^{-\lambda(q^2+\mu^2)},
\end{equation}
and the integral over $\mathbf q$ in $\sigma(r)$ becomes Gaussian. This is, in essence, Schwinger's proper-time parametrization~\cite{Schwinger1951},
commonly used in quantum field theory. Performing the Gaussian integral and changing variables to $\tau=1/(4\lambda)$ we obtain, for
each $s$,
\begin{equation}\label{eq:laplace}
1-\mu(s) r\,K_1(\mu(s) r)=\int_0^\infty\!\dd\tau\,\rho_s(\tau)\,\big(1-e^{-\tau r^2}\big),
\end{equation}
with
\begin{equation}\label{eq:rho}
\rho_s(\tau)=\frac{\mu(s)^2}{4\tau^2}\,e^{-\mu(s)^2/(4\tau)}.
\end{equation}
Each term $e^{-\tau r^2}$ is the dipole factor of a Gaussian scattering center, which transfers a transverse momentum distributed as
$e^{-q^2/(4\tau)}$, with $\langle q^2\rangle=4\tau$. We refer to $\tau$ as the transverse-momentum width of the center.
The screening mass only sets the scale of the widths: with $\varepsilon=\mu^2/(4\tau)$, $\rho_s(\tau)\,\dd\tau=e^{-\varepsilon}\dd\varepsilon$.

Let us analyze the result. $\rho_s$ is a probability density, $\int_0^\infty\rho_s\,\dd\tau=1$, but it has a heavy tail,
$\rho_s\simeq\mu^2/(4\tau^2)$ for $\tau\to\infty$, so that the moments $\int\rho_s\,\tau^m\,\dd\tau$ diverge for $m\ge1$. This tail is
the Coulomb tail $1/q^4$ of $V(q)$, that is, the hard kicks. It will be relevant in the Monte Carlo of Sec.~\ref{sec:mc}: it is the
reason why we integrate exactly over the width of a center whenever possible (Sec.~\ref{sec:mc1}), and why we introduce importance
sampling at short times (Sec.~\ref{sec:umin}).

\subsection{The medium as a Poisson process}\label{sec:poisson}
Beyond the apparent simplification of writing the potential as an average of Gaussians, the representation~\eqref{eq:laplace}
allows us to interpret the interaction with the medium as a Poisson process. We first discuss an identity valid for Poisson processes in
general, and then apply it to our problem.

Consider points $(s_j,\tau_j)$ in the plane, generated as follows: we divide the plane into cells $\dd s\,\dd\tau$ and, independently for
each cell, place a point in it with probability $n(s)\rho_s(\tau)\,\dd s\,\dd\tau$. This is a Poisson process in the plane with intensity
$n(s)\rho_s(\tau)$. Equivalently, the positions $s_j$ are distributed along $[0,L]$ with density $n(s)$, and each center draws its
width $\tau_j$ from $\rho_{s_j}$, independently of everything else (marking theorem~\cite[Theorem~5.6]{LastPenrose2017}). The number of centers is Poisson distributed with mean
$\chi=\int_0^Ln\,\dd s$, the opacity. We denote by $\E$ the expectation over this process.

\textit{Generating functional.} For a bounded function $f(s,\tau)$,
\begin{equation}\label{eq:gen}
\E\Big[\prod_j f(s_j,\tau_j)\Big]=e^{-\int_0^L\!\dd s\,n(s)\int_0^\infty\!\dd\tau\,\rho_s(\tau)\,[1-f(s,\tau)]}.
\end{equation}
\textit{Proof.} The cells are independent, so the expectation of the product is the product of the expectations. In a cell the factor
equals $f(s,\tau)$ with probability $n\rho_s\,\dd s\,\dd\tau$ and $1$ otherwise; its expectation is
$1-n\rho_s\,\dd s\,\dd\tau\,(1-f)$. The product over cells tends to the exponential in Eq.~\eqref{eq:gen}. $\square$\\
This is the probability generating functional of a Poisson process~\cite[Eq.~(3.5) and Theorem~3.9]{LastPenrose2017};
the proof above is complete and does not require homogeneity.

In our case, in each slice $\dd s$ there is a scattering center with probability $n(s)\,\dd s$, and each center carries a Gaussian
potential whose width is drawn from $\rho_s$; the Yukawa potential emerges from the average over the widths. Indeed, with
$f(s,\tau)=e^{-\tau r(s)^2}$, Eqs.~\eqref{eq:gen} and~\eqref{eq:laplace} give, for \emph{any} path $\vrr(s)$ and any time interval $J$,
\begin{align}\label{eq:poissonv}
e^{-\int_J\dd s\,v(\vrr(s),s)}
&=\sum_{N=0}^\infty e^{-\chi_J}\frac{\chi_J^N}{N!}\prod_{i=1}^N\int_J\frac{\dd s_i\,n(s_i)}{\chi_J}\notag\\
&\qquad\times\int_0^\infty\!\dd\tau_i\,\rho_{s_i}(\tau_i)\,e^{-\tau_i\,r(s_i)^2}\notag\\
&=\E\Big[\prod_{j:\,s_j\in J}e^{-\tau_j\,r(s_j)^2}\Big]\notag\\
&=\E\Big[e^{-\int_J\dd s\sum_j\tau_j\,\delta(s-s_j)\,r(s)^2}\Big].
\end{align}
with $\chi_J=\int_J n\,\dd s$. For a fixed configuration the exponent is \emph{quadratic} in $\vrr$: a harmonic oscillator whose frequency acts only at the instants $s_j$.

Note that the number $N$ of centers in a configuration is not the order in the opacity expansion. Each power of
$v=n\,(1-\int\rho_s\,e^{-\tau r^2})$ in that expansion mixes a real and a virtual term, whereas in Eq.~\eqref{eq:poissonv} the
virtual terms are resummed in the Poisson weight $e^{-\chi}$ of each configuration.
This weight is the elastic Sudakov factor that is resummed in the resummed opacity expansion~\cite{IsaksenTakacsTywoniuk2023},
which then expands in the number of real scatterings. Here, instead, each configuration, with any number of centers, is solved in
closed form, and the average over configurations sums all of them.

\subsection{Kernel and broadening as a single functional}\label{sec:single}
Consider the complete path on $[t_1,\infty)$, equal to $\vrr(s)$ between $t_1$ and $t_2$ and equal to $\vz$ (frozen) after $t_2$. The
product of kernel and broadening factor in Eq.~\eqref{eq:start} then contains
\begin{equation}
e^{-\int_{t_1}^{t_2}\!v(\vrr(s))\,\dd s}\;e^{-\int_{t_2}^{\infty}\!v(\vz)\,\dd s}
=e^{-\int_{t_1}^{\infty}\!v(\vrr(s))\,\dd s},
\end{equation}
that is, the functional~\eqref{eq:poissonv} evaluated on a \emph{single} path. For a single configuration:
the centers with $t_1<s_j<t_2$ enter the kernel, with the varying path; the centers with $s_j>t_2$ enter the broadening factor with
$r=z$ fixed, $\prod_{s_j>t_2}e^{-\tau_jz^2}=e^{-T(t_2)\,z^2}$, with $T(t_2)=\sum_{s_j>t_2}\tau_j$; and the centers with $s_j<t_1$ do
not contribute. Exchanging $\E$ with the path integral,
\begin{gather}
e^{-\frac12\int_{t_2}^\infty n\sigma(z)}\,\nabla_{\vy}\!\cdot\nabla_{\vz}\K\notag\\
=\E\Big[e^{-T(t_2)z^2}\,\nabla_{\vy}\!\cdot\nabla_{\vz}\K_c(t_2,\vz;t_1,\vy)\Big],\label{eq:Kconf}\\
\K_c=\int\mathcal D\vrr\,\exp\Big\{\int_{t_1}^{t_2}\!\dd s\Big[\frac{i\omega}{2}\dot\vrr^{\,2}-\sum_j\tau_j\delta(s-s_j)\,r^2\Big]\Big\}.\notag
\end{gather}
Configuration by configuration, everything is Gaussian: an oscillator with an impulsive frequency.

Since centers in disjoint intervals are independent, for fixed $t_2$ the expectation would also factorize,
$\E[\K_c\,e^{-Tz^2}]=\E[\K_c]\,\E[e^{-Tz^2}]$. The advantage of using a single configuration is a different one: configuration by
configuration everything is Gaussian in $\vz$, and when $t_2$ moves a center passes continuously from one factor to the other
(Sec.~\ref{sec:t2}), which allows the integral over $t_2$ to be done in closed form.

\section{Master formula: spectrum differential in $\mathbf k$ and with a cut}\label{sec:master}
By Eqs.~\eqref{eq:start} and~\eqref{eq:Kconf}, for a fixed configuration we need
\begin{multline}\label{eq:Dc}
\mathcal D_c(t_1,t_2;\vk)\equiv\int\!\dd^2z\;e^{-i\vk\cdot\vz}\,e^{-T(t_2)z^2}\\
\times\nabla_{\vy}\!\cdot\nabla_{\vz}\,\K_c(t_2,\vz;t_1,\vy)\Big|_{\vy=0}.
\end{multline}
Everything in this section refers to a fixed configuration, with arbitrary $n(s)$ and $\mu(s)$, which enter only through the $s_j$ and
the $\tau_j$; the expectation $\E$ is taken at the end.

\subsection{Center-by-center transverse integrals}\label{sec:discrete}
For a fixed configuration, the integrand of Eq.~\eqref{eq:Dc} is a product of three types of factors: free Fresnel kernels between
consecutive instants, $\frac{\omega}{2\pi i\Delta}\exp\{\frac{i\omega(\vrr'-\vrr)^2}{2\Delta}\}$; at each center $t_1<s_j<t_2$, the factor
$e^{-\tau_jr_j^2}$, with $\vrr_j$ the transverse position at $s_j$; and, at $t_2$, the broadening factor $e^{-T(t_2)z^2}$ together with
the phase $e^{-i\vk\cdot\vz}$. We integrate over the positions $\vrr_j$ one at a time, in time order.

\textit{Accumulated kernel.} After integrating over all positions before an instant $t$, the result as a function of the position
$\vrr$ at $t$ is
\begin{equation}\label{eq:kacum}
\mathcal N(t)\,\exp\Big\{\frac{i\omega}{2}\big[p(t)\,r^2-2q(t)\,\vrr\cdot\vy\big]+O(y^2)\Big\},
\end{equation}
where the term in $y^2$ is not needed, since only $\nabla_{\vy}$ at $\vy=0$ enters. The family~\eqref{eq:kacum} is closed under the
two operations that appear, with the rules of Table~\ref{tab:rules}. A free segment is the two-dimensional Gaussian integral
$\int\dd^2r\,\frac{\omega}{2\pi i\Delta}e^{\frac{i\omega}{2\Delta}(\vrr'-\vrr)^2}\,e^{\frac{i\omega}{2}[pr^2-2q\vrr\cdot\vy]}$; with
$\zeta\equiv1/p$, its rule for $p$ is simply $\zeta\to\zeta+\Delta$. A center multiplies by $e^{-\tau_jr^2}=\exp\{\frac{i\omega}{2}a_jr^2\}$, with
\begin{equation}\label{eq:aj}
a_j=\frac{2i\tau_j}{\omega}.
\end{equation}
\begin{table}[htb]
\caption{Rules for the accumulated kernel~\eqref{eq:kacum}.}\label{tab:rules}
\begin{ruledtabular}
\begin{tabular}{lccc}
step & $p$ & $q$ & $\mathcal N$\\
\hline
first segment, length $\Delta$ from $(t_1,\vy)$ & $1/\Delta$ & $1/\Delta$ & $\omega/(2\pi i\Delta)$\\
free segment, length $\Delta$ & $\frac{p}{1+p\Delta}$ & $\frac{q}{1+p\Delta}$ & $\frac{\mathcal N}{1+p\Delta}$\\
center $j$ & $p+a_j$ & $q$ & $\mathcal N$\\
\end{tabular}
\end{ruledtabular}
\end{table}

The Fresnel integrals with complex parameters have a precise distributional meaning. Here, moreover, they are absolutely convergent in
almost all cases: each intermediate position $\vrr_j$ sits at a center and carries the real damping $e^{-\tau_jr_j^2}$, and the final
position $\vz$ carries $e^{-Tz^2}$ with $T>0$ as soon as there is a center after $t_2$. Only for $T=0$ is the integral over $\vz$ a pure
Fresnel integral, to be understood in the distributional sense.

\textit{The function $S$.} We define
\begin{equation}\label{eq:pqS}
S\equiv\frac1q,\qquad \Sd\equiv\frac pq,\qquad\text{so that}\quad \mathcal N=\frac{\omega}{2\pi i\,S}.
\end{equation}
Between centers $\zeta=1/p=S/\Sd$ grows as $\zeta+\Delta$, so that $S$ is linear in $t$ with constant slope $\Sd$. At a center, $p\to p+a_j$
with $q$ fixed means
\begin{equation}\label{eq:salto}
\Sd_j^{+}=\Sd_j^{-}+a_j\,S_j,\qquad S_j\equiv S(s_j;t_1),
\end{equation}
with $S$ continuous. Without medium, $S=t_2-t_1$; centers with $s_j<t_1$ do not enter. Expanding the recursion, $S$ is a
\emph{multilinear} polynomial in the $a_j$,
\begin{equation}\label{eq:continuante}
S(t_2;t_1)=\sum_{U}\ \prod_{i\in U}a_i\ \cdot\ (s_{u_1}-t_1)(s_{u_2}-s_{u_1})\cdots(t_2-s_{u_k}),
\end{equation}
where the sum runs over the subsets $U=\{u_1<\dots<u_k\}$ of the centers in $(t_1,t_2)$, the term $U=\varnothing$ being $t_2-t_1$.
This is the Gelfand--Yaglom structure~\cite{GelfandYaglom1960} that reappears in Sec.~\ref{sec:gy}.

\textit{Broadening as one more kick.} Multiplying the kernel at $t_2$ by $e^{-T(t_2)z^2}$ amounts, in Table~\ref{tab:rules}, to
$p\to p+2iT/\omega$: the accumulated effect of all centers after $t_2$, applied at once, with no free segments between them. The total
coefficient of $-z^2$ in the exponent is then
\begin{equation}\label{eq:thetap}
\vartheta\equiv T-\eta=-\frac{i\omega}{2}\Big(p+\frac{2iT}{\omega}\Big),\qquad \eta\equiv\frac{i\omega p}{2}=\frac{i\omega\,\Sd}{2S}.
\end{equation}
When $t_2$ crosses a center, its kick passes from ``after $t_2$'' to ``before $t_2$'' without changing the sum of kicks applied at that
instant: $\vartheta$ is continuous by construction. This is the transparent form of the statement that kernel and broadening are a
single functional (Sec.~\ref{sec:single}).

\textit{Positivity of $\Re\vartheta$.}
From Eq.~\eqref{eq:thetap}, $\Re\vartheta=T+\frac{\omega}{2}\Im p$. On the first segment $p=1/\Delta$ is real. At each center
$\Im p$ increases by $2\tau_j/\omega>0$. On a free segment $\zeta\to\zeta+\Delta$ with $\Delta$ real preserves the sign of $\Im\zeta$, and
$\Im p=-\Im\zeta/|\zeta|^2$, so $\Im p\ge0$ is preserved. Hence $\Re\vartheta\ge T\ge0$, with strict inequality as soon as there is any
center after $t_1$.

\textit{Derivative at $\vy=0$.} From Eqs.~\eqref{eq:kacum} and~\eqref{eq:pqS},
$\nabla_{\vy}\K_c|_{\vy=0}=-i\omega q\,\vz\,\mathcal N e^{\eta z^2}=-\frac{\omega^2}{2\pi S^2}\,\vz\,e^{\eta z^2}$. Applying
$\nabla_{\vz}\cdot$, with $\nabla_{\vz}\cdot\vz=2$ in two dimensions,
\begin{equation}\label{eq:dd}
\nabla_{\vy}\!\cdot\nabla_{\vz}\,\K_c\big|_{\vy=0}=-\frac{\omega^2}{\pi S^2}\,\big(1+\eta z^2\big)\,e^{\eta z^2}.
\end{equation}

\subsection{Integrals over $\mathbf z$ and over $|\mathbf k|<K$}\label{sec:zk}
By Eq.~\eqref{eq:dd}, the integrand of Eq.~\eqref{eq:Dc} is $e^{-i\vk\cdot\vz}\,(1+\eta z^2)\,e^{-\vartheta z^2}$. Since
$\Re\vartheta>0$, these are two Gaussian integrals in two dimensions,
\begin{gather*}
\int\!\dd^2z\,e^{-i\vk\cdot\vz-\vartheta z^2}=\frac{\pi}{\vartheta}\,e^{-k^2/4\vartheta},\\
\int\!\dd^2z\,z^2\,e^{-i\vk\cdot\vz-\vartheta z^2}=\frac{\pi}{\vartheta^2}\Big(1-\frac{k^2}{4\vartheta}\Big)e^{-k^2/4\vartheta}.
\end{gather*}
Thus, per configuration and for fixed $t_1$, $t_2$,
\begin{equation}\label{eq:difconf}
\mathcal D_c(t_1,t_2;\vk)
=-\frac{\omega^2}{S^2\vartheta}\,e^{-k^2/4\vartheta}\Big[1+\frac{\eta}{\vartheta}\Big(1-\frac{k^2}{4\vartheta}\Big)\Big].
\end{equation}
For the integral over $|\vk|<K$ we use $\dd^2k=\pi\,\dd(k^2)$ and $Q\equiv e^{-K^2/4\vartheta}$. The terms in $\eta/\vartheta$ without
$K$ cancel, and
\begin{multline}
\int_{|\vk|<K}\!\dd^2k\;\frac{\pi}{\vartheta}e^{-k^2/4\vartheta}\Big[1+\frac{\eta}{\vartheta}\Big(1-\frac{k^2}{4\vartheta}\Big)\Big]\\
=4\pi^2\Big[1-Q\Big(1-\frac{\eta K^2}{4\vartheta^2}\Big)\Big].
\end{multline}
With the prefactor of Eq.~\eqref{eq:start} and $\I_K\equiv\frac{\pi}{2\alpha_sC_R}\,\omega\frac{\dd I}{\dd\omega}\big|_{k<K}$,
\begin{gather}
\I_K=\Re\,\E\Big[\int\!\dd t_1\!\int_{t_1}^\infty\!\dd t_2\;\big(-h(t_2)\big)\Big]-(\text{vacuum}),\label{eq:IKdoble}\\
h(t_2)\equiv\frac{1}{S^2}\Big[1-Q+\frac{\eta\,Q\,K^2}{4\vartheta^2}\Big].\notag
\end{gather}

\subsection{Integral over $t_2$}\label{sec:t2}
For fixed $t_1$ and configuration, $S$, $\Sd$, $T$, $\eta$, $\vartheta$ and $Q$ are functions of $t_2$.

\textit{(i) Primitive between centers.} Between two centers $\Sd$ and $T$ are constant. Let
\begin{equation}\label{eq:P}
P(t_2)\equiv\frac{Q-1}{S\,\Sd}.
\end{equation}
Then $\partial_{t_2}(1/S\Sd)=-1/S^2$; $\dot\eta=-\eta\Sd/S$, hence $\dot\vartheta=-\dot\eta=\eta\Sd/S$; and
$\dot Q=Q\,\frac{K^2}{4\vartheta^2}\,\dot\vartheta$. Adding up,
\begin{equation}
\partial_{t_2}P=\frac{1-Q}{S^2}+\frac{\dot Q}{S\Sd}=h(t_2).
\end{equation}

\textit{(ii) Continuity at the centers.} $S$ and $\vartheta$ are continuous (Sec.~\ref{sec:discrete}), hence so is $Q$; only $\Sd$
jumps. At $t_2=s_j$,
\begin{equation}\label{eq:theta}
\vartheta_j\equiv T_j^+-\frac{i\omega\,\Sd_j^+}{2S_j}=T_j^- -\frac{i\omega\,\Sd_j^-}{2S_j},
\end{equation}
with $T_j^+=\sum_{i>j}\tau_i$ and $T_j^-=\tau_j+T_j^+$: the jump of $T$ exactly compensates that of $\Sd$. The value of $T$ at
$t_2=s_j$ itself is ambiguous, but it is a set of measure zero, the piecewise sum only uses the one-sided limits, and the combination
that enters is continuous.

\textit{(iii) Piecewise sum.} $\int_{t_1}^\infty h\,\dd t_2=\sum_{\text{segments}}[P(\text{end})-P(\text{start})]$, with three types of
contribution.
(a) \emph{Endpoint $t_2\to t_1^+$.} $S\to0$, $\Sd=1$, $\eta\simeq i\omega/(2S)\to\infty$, $\vartheta\simeq-\eta$, $K^2/4\vartheta\to0$ and
  $P\to K^2/(2i\omega)$: a purely imaginary constant, independent of the configuration. It is exactly the vacuum contribution, which
  cancels in the subtraction.
(b) \emph{Endpoint $t_2\to\infty$.} $S$ grows linearly and $|Q|\le1$, so $P\to0$.
(c) \emph{Center $s_j>t_1$.} With $S$ and $Q$ continuous and the jump~\eqref{eq:salto},
\[
P(s_j^-)-P(s_j^+)=\frac{Q_j-1}{S_j}\Big(\frac1{\Sd_j^-}-\frac1{\Sd_j^+}\Big)=\frac{(Q_j-1)\,a_j}{\Sd_j^-\,\Sd_j^+},
\]
with $Q_j=e^{-K^2/4\vartheta_j}$.
For $t_1>L$ there are no later centers and the contribution vanishes. We arrive at the master formula, valid for arbitrary $n(s)$ and
$\mu(s)$:
\begin{multline}\label{eq:IK}
\I_K=\Re\,\E\bigg[\int_0^L\!\dd t_1\\
\times\sum_{j:\,s_j>t_1}\frac{a_j\big(1-e^{-K^2/4\vartheta_j}\big)}{\Sd_j^-\,\Sd_j^+}\bigg],
\end{multline}
with $\vartheta_j$ given by Eq.~\eqref{eq:theta}, $S_j=S(s_j;t_1)$, $\Sd_j^\mp$ the slopes just before and after center $j$, $a_j=2i\tau_j/\omega$,
and $\E$ the expectation over the Poisson process of Sec.~\ref{sec:poisson}. Differentiating with respect to $K^2$,
\begin{multline}\label{eq:dif}
\omega\frac{\dd I}{\dd\omega\,\dd k^2}=\frac{2\alpha_sC_R}{\pi}\\
\times\Re\,\E\Big[\int_0^L\!\dd t_1\sum_{j:\,s_j>t_1}\frac{a_j}{4\vartheta_j\,\Sd_j^-\Sd_j^+}\,e^{-k^2/4\vartheta_j}\Big],
\end{multline}
a mixture of complex Gaussians in $k$.

\textit{Reading.} The gluon is emitted at $t_1$; the centers between $t_1$ and $s_j$ build $S$ through the jumps~\eqref{eq:salto}; then
comes a ``special'' center $j$; the centers after $j$ only count through $T_j^+$ inside $\vartheta_j$, that is, through the broadening of
the already emitted gluon. The profile of the medium only enters through where the centers fall ($n$) and through their widths ($\mu$).

\section{Spectrum integrated over all $\mathbf k$}\label{sec:integrated}
Integrating Eq.~\eqref{eq:start} over all $\vk$ gives $\int\dd^2k\,e^{-i\vk\cdot\vz}=(2\pi)^2\delta^{(2)}(\vz)$, and since $\sigma(0)=0$ the
broadening factor disappears. With $\I\equiv\frac{\pi}{2\alpha_sC_R}\,\omega\frac{\dd I}{\dd\omega}$ and Eq.~\eqref{eq:dd} at $\vz=0$,
\begin{equation}\label{eq:Iint}
\I=\Re\,\E\Big[-\int_0^\infty\!\dd t_1\!\int_{t_1}^\infty\!\dd t_2\Big(\frac{1}{S(t_2;t_1)^2}-\frac{1}{(t_2-t_1)^2}\Big)\Big],
\end{equation}
where the vacuum has $S_0=t_2-t_1$. This is the structure of Eqs.~(3.7)--(3.8) of Arnold \cite{Arnold2009}. It is also the $K\to\infty$ limit of the
master formula: with $Q_j\to0$, Eq.~\eqref{eq:IK} gives
\begin{equation}\label{eq:Kinf}
\I=\Re\,\E\Big[\int_0^L\!\dd t_1\sum_{j:\,s_j>t_1}\frac{a_j}{\Sd_j^-\Sd_j^+}\Big].
\end{equation}
The next two subsections bring this spectrum to the form $\I=\E\ln|\Phi|$.

\subsection{Arnold's route: the Wronskian identity}\label{sec:arnold}
\textit{Integral over $t_2$}~\cite[Eqs.~(3.9)--(3.15)]{Arnold2009}. $S$ and $C=-\partial_{t_1}S$ solve the same equation in $t_2$, with constant
Wronskian $C\Sd-S\,\partial_{t_2}C=1$, evaluated at $t_2=t_1$, where $S=0$ and $\Sd=C=1$. Dividing by $S^2$,
$\partial_{t_2}(C/S)=-1/S^2$, and for the vacuum $C_0/S_0=1/(t_2-t_1)$. At $t_2\to t_1^+$, before the first center, $S=t_2-t_1$ and $C=1$
exactly, so the lower limit does not contribute. For $t_2\ge L$, $S$ is a straight line, $S=S(L;t_1)+\varphi(t_1)(t_2-L)$ with
$\varphi(t_1)\equiv\Sd(t_2;t_1)|_{t_2\ge L}$, and $C/S\to-\varphi'(t_1)/\varphi(t_1)$. Hence
\begin{equation}\label{eq:t2int}
\int_{t_1}^{\infty}\!\dd t_2\Big(\frac1{S^2}-\frac1{(t_2-t_1)^2}\Big)=\frac{\varphi'(t_1)}{\varphi(t_1)} .
\end{equation}

\textit{Integral over $t_1$}~\cite[Eqs.~(3.19)--(3.20)]{Arnold2009}. For $t_1\ge L$ there are no centers after $t_1$ and $\varphi=1$. Then
$-\int_0^\infty\dd t_1\,\varphi'/\varphi=\ln\varphi(0)$, and with $\Phi\equiv\varphi(0)=f'(L)$, $f(s)\equiv S(s;0)$,
\begin{equation}\label{eq:Ilog}
\I=\Re\,\E\big[\ln\Phi\big]=\E\big[\ln|\Phi|\big],\qquad \Phi=f'(L).
\end{equation}
The real part removes the branch ambiguity of the logarithm.

\textit{Link with the master formula.} Doing the same $t_2$ integral with the primitive of Sec.~\ref{sec:t2} at $K\to\infty$, where
$P\to-1/(S\Sd)$, the left-hand side of Eq.~\eqref{eq:t2int} equals $-\sum_{s_j>t_1}a_j/(\Sd_j^-\Sd_j^+)$. Therefore
\begin{equation}
\sum_{j:\,s_j>t_1}\frac{a_j}{\Sd_j^-\Sd_j^+}=-\frac{\varphi'(t_1)}{\varphi(t_1)},
\end{equation}
and Eqs.~\eqref{eq:Kinf} and~\eqref{eq:Ilog} coincide.
Arnold's paper~\cite{Arnold2009} treats precisely an arbitrary $\hat q(t)$, i.e., expanding media, in the HO approximation; here the same identity is
applied configuration by configuration.

\subsection{Gelfand--Yaglom route: $\Phi$ as a multilinear determinant}\label{sec:gy}
\textit{Lemma.} $\Phi=f'(L)=\det(\mathbf 1+\mathbf A G)$, with $\mathbf A=\mathrm{diag}(a_1,\dots,a_N)$ and $G_{ij}=\min(s_i,s_j)$.

\textit{Proof.} \emph{Multilinearity.} By the rules of Table~\ref{tab:rules}, or with the transfer matrices (free segment
$(f,f')\mapsto(f+\Delta f',f')$; center $j$, $(f,f')\mapsto(f,f'+a_jf)$), each center acts linearly in its $a_j$, so that $f'(L)$ is
multilinear, $f'(L)=\sum_{U}c_U\prod_{i\in U}a_i$, with the sum over subsets $U$ of centers.
\emph{Coefficients.} Let $U=\{u_1<\dots<u_k\}$. The coefficient $c_U$ can be computed with only the centers in $U$. After the last one,
$f'(L)=f'(u_k^-)+a_kf(u_k)$, and $f(u_k)=f(u_{k-1})+(u_k-u_{k-1})f'(u_{k-1}^+)$. Since $f(u_{k-1})$ does not depend on $a_{k-1}$, the
coefficient of $a_1\cdots a_{k-1}$ in $f(u_k)$ is $(u_k-u_{k-1})$ times that of $f'(u_{k-1}^+)$, which is the problem with $k-1$ centers.
With $c_{\{u_1\}}=u_1$, $c_U=u_1(u_2-u_1)\cdots(u_k-u_{k-1})$.
\emph{Determinant.} Since $\mathbf A$ is diagonal, $\det(\mathbf 1+\mathbf AG)=\sum_U\prod_{i\in U}a_i\,\det G_U$, with $G_U$ the
principal minors. Subtracting from each row of $G_U$ the previous one gives a triangular matrix with diagonal $u_1,u_2-u_1,\dots$, so
$\det G_U=c_U$. $\square$

This is the analogue, for the final slope, of the multilinear polynomial~\eqref{eq:continuante}. In practice $\Phi$ is computed with the
M\"obius recursion: with $\zeta=f/f'$ (the $1/p$ of Sec.~\ref{sec:discrete} with $t_1=0$), $\zeta\to\zeta+\Delta$ between centers and
$f'\to f'(1+a_j\zeta)$ at each center, so that $\zeta\to\zeta/(1+a_j\zeta)$ and
\begin{equation}\label{eq:mobius}
\Phi=\prod_{j=1}^N\big(1+a_j\,\zeta_j^-\big),\qquad \zeta(0)=0,
\end{equation}
where $\zeta_j^-$ is the value just before center $j$.

\subsection{One center: closed-form average over $\tau$}\label{sec:V}
Each factor of Eq.~\eqref{eq:mobius} only depends on the width of its own center and on the earlier centers. For the center at $s$,
$a\zeta=2i\tau\zeta/\omega=\beta/\varepsilon$, with
\begin{equation}
\beta\equiv\frac{i\mu(s)^2\zeta}{2\omega}
\end{equation}
and $\varepsilon\sim\mathrm{Exp}(1)$ (Sec.~\ref{sec:laplace}). The average over the width is then elementary,
\begin{equation}\label{eq:V}
\int_0^\infty\!\dd\varepsilon\,e^{-\varepsilon}\ln\Big(1+\frac\beta\varepsilon\Big)=V(\beta)\equiv\ln\beta+\gamma_E+e^{\beta}E_1(\beta),
\end{equation}
since $\int e^{-\varepsilon}\ln(\varepsilon+\beta)\,\dd\varepsilon=\ln\beta+e^\beta E_1(\beta)$ and
$\int e^{-\varepsilon}\ln\varepsilon\,\dd\varepsilon=-\gamma_E$.
This is the piece used by the Monte Carlo of the integrated spectrum (Sec.~\ref{sec:mc1}). With a single center ($\zeta=s$) it gives the
first order in opacity, $\int_0^L\dd s\,n(s)\,\Re V\big(i\mu(s)^2s/2\omega\big)$, which for a static medium is the GLV result,
Eq.~(B.4) of Ref.~\cite{AndresApolinarioDominguez2020} in the limit $\bar R\to\infty$ (no kinematic cut; see Sec.~\ref{sec:opacity}).

\section{Monte Carlo}\label{sec:mc}
Equations~\eqref{eq:IK}, \eqref{eq:dif} and~\eqref{eq:Ilog} are averages over configurations of the Poisson process. The Monte Carlo
method is the standard one~\cite{James1980}: independent samples $X_1,\dots,X_n$ of an estimator $X$ whose expectation is the quantity sought, and the
result $\bar X\pm s_X/\sqrt n$. What is not trivial is the choice of $X$. We use four variance-reduction techniques, all of general use:
(i) a marked center (Mecke's formula), which replaces a sum over centers by an integral over the position of one ``special'' center;
(ii) exact integration of one variable, the width of the marked center, whenever possible;
(iii) importance sampling of the time between the emission and the marked center; and
(iv) a control variate, the estimator without cut, whose mean is the integrated spectrum.
Everything is written for arbitrary $n(s)$ and $\mu(s)$; the static medium is a particular case.

\subsection{Mecke's formula}\label{sec:mecke}
For a function $h(x;\mathcal C)$ of a point $x=(s,\tau)$ and of the configuration $\mathcal C$ of the Poisson process of
Sec.~\ref{sec:poisson},
\begin{multline}\label{eq:mecke}
\E\Big[\sum_{x_j\in\mathcal C}h(x_j;\mathcal C)\Big]\\
=\int_0^L\!\dd s\,n(s)\int_0^\infty\!\dd\tau\,\rho_s(\tau)\;\E\big[h\big((s,\tau);\mathcal C\cup\{(s,\tau)\}\big)\big].
\end{multline}
This is Mecke's formula \cite{Mecke1967}, \cite[Theorem~4.1]{LastPenrose2017}. With the cells of Sec.~\ref{sec:poisson} it follows in a few lines: the cell at $(s,\tau)$ contributes to
the sum only if it contains a center, which happens with probability $n\rho_s\,\dd s\,\dd\tau$; by the independence between cells, the
rest of the configuration has the same law as if that cell did not exist, and an infinitesimal cell does not change that law. $\square$\\
In practice one samples the position $s$ of the marked center with density $n$, and only those centers that influence its term. Each
sample contributes a single term instead of a sum, and centers that do not influence it need not be sampled.

\subsection{Dimensionless variables and sampling in the opacity coordinate}\label{sec:opacity}
We measure times in units of $L$ and transverse momenta in units of a reference screening mass $\mu_{\rm ref}$:
\begin{gather}
\hat s=\frac sL,\qquad \chi(\hat s)=n(\hat sL)\,L,\qquad m_2(\hat s)=\frac{\mu(\hat sL)^2}{\mu_{\rm ref}^2},\notag\\
g_{\rm ref}=\frac{i\mu_{\rm ref}^2L}{2\omega},\qquad
\alpha_j=a_jL=\frac{g_{\rm ref}\,m_2(\hat s_j)}{\varepsilon_j},\notag\\
\Theta_j=\sum_{i>j}\frac{m_2(\hat s_i)}{\varepsilon_i},\qquad \kappa=\frac{K}{\mu_{\rm ref}},
\end{gather}
with $\varepsilon_j\sim\mathrm{Exp}(1)$ (Sec.~\ref{sec:laplace}). Then $K^2/4\vartheta_j=\kappa^2/D_j$ with
\begin{equation}\label{eq:Dj}
D_j=\Theta_j+\frac{\Sd_j^+}{g_{\rm ref}\hat S_j},\qquad \hat S_j=S_j/L .
\end{equation}
For a static medium $\chi$ is the opacity, $m_2=1$ and $g_{\rm ref}=i\gamma$ with $\gamma=\mu^2L/(2\omega)=1/x$, where, as in
Ref.~\cite{AndresApolinarioDominguez2020}, $x=\omega/\bar\omega_c$ and $\bar\omega_c=\mu^2L/2$. The kinematic cut $K=\omega$ then corresponds to
$\kappa^2=\bar Rx^2/2$, with $\bar R=\bar\omega_cL=\mu^2L^2/2$; $\bar R\to\infty$ means no cut.

\textit{Sampling.} With the opacity coordinate $X(\hat s)=\int_0^{\hat s}\chi$, the probability of a center in a slice is
$\chi\,\dd\hat s=\dd X$: in the variable $X$ the process is homogeneous with unit rate on $[0,\chi]$, where $\chi\equiv X(1)$ is the total
opacity. We draw points uniformly in $X$, their number being Poisson distributed with mean equal to the length of the interval in $X$, and
map them back with $\hat s=X^{-1}(X)$. Stretches without medium ($n=0$) receive no centers. In Mecke's formula, $\int n\,\dd s=\int\dd X$.

\subsection{Case 1: spectrum integrated over $\mathbf k$}\label{sec:mc1}
By Eq.~\eqref{eq:mobius}, $\ln\Phi=\sum_j\ln(1+a_j\zeta_j^-)$, where $\zeta_j^-$ only depends on the centers \emph{before} $s_j$. By
Eqs.~\eqref{eq:mecke} and~\eqref{eq:V}, with the width of the marked center integrated exactly,
\begin{equation}\label{eq:mc1}
\I=\int_0^{\chi}\!\dd X_*\;\E\Big[\Re V\big(g_{\rm ref}\,m_2(\hat s_*)\,\hat\zeta^-(\hat s_*)\big)\Big],
\end{equation}
with $\hat s_*=X^{-1}(X_*)$, and where $\hat\zeta^-(\hat s_*)=\zeta/L$ just before $\hat s_*$ is generated by the centers in $(0,\hat s_*)$. One sample consists of:
(a) $X_*$ uniform in $[0,\chi]$ (stratified in the code) and $\hat s_*=X^{-1}(X_*)$; (b) $\mathrm{Poisson}(X_*)$ centers in
$(0,\hat s_*)$, uniform in $X$, each with $\varepsilon_i\sim\mathrm{Exp}(1)$; (c) the M\"obius recursion $\hat\zeta\to\hat\zeta+\Delta$
between centers and $\hat\zeta\to\hat\zeta/(1+\alpha_i\hat\zeta)$ at each center, from $\hat\zeta(0)=0$ up to $\hat s_*$; and
(d) $X=\chi\,\Re V\big(g_{\rm ref}m_2(\hat s_*)\hat\zeta\big)$. Centers after $\hat s_*$ are not sampled. The exact integration over the
width of the marked center removes the largest source of variance, the Coulomb tail ($\varepsilon\to0$). As a further control variate
we use the same estimator without earlier centers, i.e., the first order in opacity, whose mean $\int_0^1\dd\hat s\,\chi\,\Re V(g_{\rm
ref}m_2\hat s)$ is computed by quadrature.

\subsection{Cases 2 and 3: with a cut and differential}\label{sec:mc23}
In Eq.~\eqref{eq:IK} there remain the integral over $t_1$ and the sum over the marked center $j$. Applying Eq.~\eqref{eq:mecke} to $j$
and passing to units of $L$, $a_j\,\dd t_1\,n(s_j)\,\dd s_j=\alpha_j\,\chi(\hat s_j)\,\dd\hat t_1\,\dd\hat s_j$, so that
\begin{multline}\label{eq:mc2}
\I_K=\Re\int_0^1\!\dd\hat t_1\int_{\hat t_1}^1\!\dd\hat s_j\;\chi(\hat s_j)\\
\times\E\Big[\frac{\alpha_j}{\Sd_j^-\Sd_j^+}\big(1-e^{-\kappa^2/D_j}\big)\Big],\qquad \Sd_j^+=\Sd_j^-+\alpha_j\hat S_j .
\end{multline}
With $u=\hat s_j-\hat t_1$, $\int_0^1\dd\hat t_1\int_{\hat t_1}^1\dd\hat s_j=\int_0^1\dd u\int_0^{1-u}\dd\hat t_1$. One sample consists of:
(a) $u$ log-uniform in $[u_{\min},1]$ and $\hat t_1$ uniform in $[0,1-u]$, with $\hat s_j=\hat t_1+u$ and weight
$w=\chi(\hat s_j)\,(1-u)\,u\,\ln(1/u_{\min})$;
(b) the centers between the emission and the marked center, $\mathrm{Poisson}\big(X(\hat s_j)-X(\hat t_1)\big)$, uniform in $X$, with their
$\varepsilon_i$; starting from $(\hat S,\Sd)=(0,1)$ at $\hat t_1$, they are propagated with the free stretches and the jumps~\eqref{eq:salto},
which gives $\hat S_j$, $\Sd_j^-$ and, with the $\varepsilon_j$ of the marked center, $\Sd_j^+$;
(c) the centers after the marked center, $\mathrm{Poisson}\big(\chi-X(\hat s_j)\big)$, uniform in $X$, which only enter through $\Theta_j$;
(d) the centers before the emission, which do not contribute and are not sampled; and
(e) the estimators
\begin{align}\label{eq:estimators}
X_K&=\Re\Big[w\,\frac{\alpha_j\,\big(1-e^{-\kappa^2/D_j}\big)}{\Sd_j^-\Sd_j^+}\Big],\notag\\
X'_{\kappa^2}&=\Re\Big[w\,\frac{\alpha_j}{\Sd_j^-\Sd_j^+}\,\frac{e^{-\kappa^2/D_j}}{D_j}\Big],
\end{align}
with $\E X_K=\I_K$ and $\E X'_{\kappa^2}=\dd\I_K/\dd\kappa^2$, so that
$\omega\frac{\dd I}{\dd\omega\,\dd\kappa^2}=\frac{2\alpha_sC_R}{\pi}\E X'_{\kappa^2}$.
Here the width of the marked center cannot be integrated in closed form, since it enters $\Sd_j^+$ and, through it, $D_j$ inside the
exponential; it is sampled.

\subsection{Log-uniform sampling in $u$ and truncation error}\label{sec:umin}
If $\varepsilon_j$ is small, $|\alpha_j|$ is large and, without intermediate centers,
$\alpha_j/(\Sd^-\Sd^+)=\alpha_j/(1+\alpha_ju)\sim1/u$ over a wide range of $u$. With $u$ uniform the estimator would be unbounded; with
the weight $u$ of the log-uniform sampling it is bounded. The log-uniform sampling needs a lower limit, and the region $u<u_{\min}$ is
omitted, with
\begin{equation}
u_{\min}=\frac{10^{-9}}{1+\kappa_{\max}^2|g_{\rm ref}|\max m_2}.
\end{equation}
This is a bias, not a statistical error, and we quantify it.

\textit{Physical argument.} The region $u\lesssim u_{\min}$ only contributes when $|\alpha_j|u\gtrsim1$, i.e., for a tiny $\varepsilon_j$: a
center with a huge width $\tau_j=\mu^2/4\varepsilon_j$, which gives a transverse kick $q^2\sim4\tau_j$. The condition $|\alpha_j|u\sim1$ states
that $uL$ is the formation time of a gluon with that momentum, $uL\sim2\omega/q^2$. At the boundary $k^2\sim q^2\sim2\omega/(u_{\min}L)$, that
is, $k^2/\omega^2\sim2/(\omega Lu_{\min})$, which is of order $10^9$ for the parameters below. Such a kick is so hard that the gluon comes out
with $k\gg\omega$, outside the kinematic cut $k\le\omega$ and outside the soft-gluon approximation in which the whole formula is written.

\textit{Size.} At first order in opacity (a single center, the marked one) the omitted part can be computed exactly by quadrature. For a
static medium with the kinematic cut $k\le\omega$, the omitted fraction ranges from $-10^{-15}$ to $-10^{-22}$ for $x=\omega/\bar\omega_c$ between
$0.01$ and $10$ and $\bar R=10^3$--$4\cdot10^4$. Without cut ($\gamma=1$, $u_{\min}=10^{-9}$) it is $3.0\cdot10^{-9}$, in agreement with the
analytic estimate $\Re V(i\gamma u_{\min})\simeq\frac\pi2\gamma u_{\min}$ divided by the first order. With several centers the order of
magnitude should be the same, since the omitted region is dominated by the hard kicks of a single center.

\subsection{Control variate, all $k$ at once, and real part}\label{sec:cv}
\textit{Control variate.} In the same sample we compute $Y=\Re[w\,\alpha_j/(\Sd_j^-\Sd_j^+)]$, the estimator without cut, whose mean is
the integrated spectrum $\I$. We take $\I$ from an independent run of the estimator of Sec.~\ref{sec:mc1}, $\hat\I\pm\delta\I$, and return
$\bar X-c(\bar Y-\hat\I)$ with $c=\mathrm{Cov}(X,Y)/\mathrm{Var}(Y)$. The variance is reduced by a factor $1-\mathrm{corr}(X,Y)^2$, and the
error of $\hat\I$ is propagated, $c^2\delta\I^2$. Since the case-1 estimator is much more precise than $\bar Y$, the control variate is
effective at little cost.

\textit{All $k$ at once.} $\kappa^2$ only enters the last factor of Eq.~\eqref{eq:estimators}, so every configuration serves a whole grid
of $\kappa^2$, and the covariance between different $\kappa^2$ is obtained as well.

\textit{Real part, sample by sample.} $\Re$ and $\E$ are linear and commute, so we take the real part of each sample before averaging.
This has two advantages: the variance of $\Re X$ does not exceed that of $X$, and the imaginary vacuum constant of the endpoint
$t_2\to t_1^+$ (Sec.~\ref{sec:t2}) disappears in each sample. What cannot be done is to take the real part \emph{inside} the computation of
a configuration, since $\Re$ does not commute with products, quotients or logarithms. The oscillator is genuinely complex,
$a_j=2i\tau_j/\omega$: $\hat S$ and $\Sd$ are propagated as complex numbers, and their phases are the LPM interference.

\subsection{Physical variables and conventions for expanding media}\label{sec:conventions}
Time is measured from the production point, and there may be vacuum before the medium; reversing the profile in time gives a different
result. Following Ref.~\cite{Andres2024}, we use $\omega=x\,\bar\omega_c$ with $\chi\,\bar\omega_c=\int\dd t\,t\,n\mu^2$, so that
$g_{\rm ref}=i\chi/(2I_1x)$ with $I_1=\int_0^1\dd\hat s\,\hat s\,\chi\,m_2$, and the kinematic cut $K=\omega$ corresponds to
$\kappa^2=\frac23\bar R\,x^2I_1^2/(\chi I_2)$ with $I_2=\int_0^1\dd\hat s\,\hat s^2\chi\,m_2$ (static medium: $\bar Rx^2/2$).

\subsection{Cost}\label{sec:cost}
The cost per sample does not depend on the profile, which only changes the Poisson intensity and $m_2$. On a laptop (two physical cores),
$10^6$ samples of the estimator of Sec.~\ref{sec:mc1} take about $3$\,s and give a relative error of $(2$--$4)\cdot10^{-4}$; $10^6$ samples of
the estimators~\eqref{eq:estimators}, for a grid of six values of $\kappa^2$, take $3$--$5$\,s, with relative errors of about $1\,\%$ for
the differential spectrum. The tasks are independent, with a single final reduction; with two processes the speed-up is $1.6$.

\section{Validation and comparison with published results}\label{sec:validation}
All numbers in this section are obtained with the publicly available code, with the statistical errors shown. The code and the
scripts that produce the figures were written with the assistance of an AI tool (see Acknowledgments) and validated as described in
this section.
The curves of
Refs.~\cite{AndresApolinarioDominguez2020,Andres2024} are read from the vector graphics of the PDF files, which gives the curves with the precision of the drawing; as a check of this reading
and of the normalization, the published first-order (GLV) curves of Ref.~\cite{AndresApolinarioDominguez2020} are reproduced to $0.9996$--$1.0002$ with
$\alpha_s=0.3$ and $C_R=4/3$.

\subsection{Internal checks}
The closed-form $t_2$ integral of Sec.~\ref{sec:t2} agrees with a direct quadrature of the integrand to $10^{-11}$; the multilinear
polynomial~\eqref{eq:continuante} and the determinant of Sec.~\ref{sec:gy} agree with the recursions to machine precision; and the
integrated spectrum obtained from the estimator of Sec.~\ref{sec:mc1} agrees with the $K\to\infty$ limit of the estimator of
Sec.~\ref{sec:mc23}, which is a different estimator of the same quantity. The integral over $\kappa^2$ of the differential estimator
reproduces the estimator with a cut sample by sample. The bias from the truncation $u_{\min}$ was discussed in Sec.~\ref{sec:umin}.

\subsection{First order in opacity}
Table~\ref{tab:N1} compares the first order in opacity of the master formula, computed by quadrature, with the closed expressions
(B.3) and (B.4) of Ref.~\cite{AndresApolinarioDominguez2020} for the differential spectrum and for the spectrum with the kinematic cut. Without cut, the first order
of Sec.~\ref{sec:V} reproduces the GLV result to twelve digits.
\begin{table}[htb]
\caption{First order in opacity (static medium, $\gamma=1/x$): master formula at $N=1$ (quadrature) versus Eqs.~(B.4) (cut,
$\kappa^2=\bar Rx^2/2$) and (B.3) (differential) of Ref.~\cite{AndresApolinarioDominguez2020}, per unit of $\frac{2\alpha_sC_R}{\pi}\chi$.}\label{tab:N1}
\begin{ruledtabular}
\begin{tabular}{lcccc}
 & $x$ & $\kappa^2$ & master formula & Ref.~\cite{AndresApolinarioDominguez2020}\\
\hline
cut, $\bar R=50$ & $0.05$ & $0.0625$ & $0.0532891$ & $0.0532891$\\
 & $1$ & $25$ & $0.2943896$ & $0.2943896$\\
 & $5$ & $625$ & $0.1201906$ & $0.1201826$\\
cut, $\bar R=1000$ & $0.05$ & $1.25$ & $0.7862587$ & $0.7862587$\\
 & $1$ & $500$ & $0.5025957$ & $0.5025957$\\
 & $5$ & $12500$ & $0.1361501$ & $0.1361425$\\
differential & $0.03$ & $0.03$ & $1.2358926$ & $1.2358926$\\
 & $0.3$ & $1$ & $0.1096906$ & $0.1096906$\\
 & $1$ & $10$ & $0.0178158$ & $0.0178158$\\
\end{tabular}
\end{ruledtabular}
\end{table}

\subsection{Comparison with the results of Andr\'es et al.}
The master formula~\eqref{eq:IK} has the same structure as Eq.~(3.3) of Ref.~\cite{AndresApolinarioDominguez2020}: the integral over $t_2$ is done analytically, one
scattering is singled out, the kernel runs from the emission to that scattering and the broadening from it onwards, and the integrand
vanishes outside the medium. Through Mecke's formula (Sec.~\ref{sec:mecke}), the sum over $j$ becomes the integral
$\int\dd s\,n(s)\int\dd\tau\,\rho_s$, the analogue of the insertion $\int\dd s\,n(s)\,\sigma$ in that equation.

\begin{figure}[htb]
\includegraphics[width=\columnwidth]{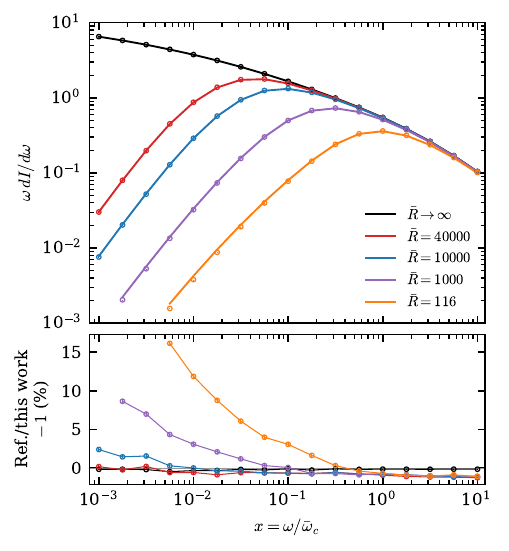}
\caption{Static medium, $\chi=5$: spectrum without cut and with the kinematic cut $k\le\omega$ for several $\bar R$. Lines: Fig.~3 of
Ref.~\cite{AndresApolinarioDominguez2020}; points: this work. Lower panel: relative difference, Ref.~\cite{AndresApolinarioDominguez2020}/this work $-\,1$, in \%. Published points below the frame of the published panel
($\omega\,dI/d\omega<10^{-3}$) are not compared. The figure was produced with scripts written with the assistance of an AI tool (see Sec.~\ref{sec:validation} and Acknowledgments).}\label{fig:static}
\end{figure}

\textit{Static medium.} Figure~\ref{fig:static} compares the spectrum for a static medium with $\chi=5$, without cut and with the
kinematic cut $k\le\omega$ for several values of $\bar R$, with Fig.~3 of Ref.~\cite{AndresApolinarioDominguez2020}. Without cut, their curve lies
$0.1$--$0.4\,\%$ below our result. With the cut, their curves lie $0.7$--$1.2\,\%$ below our results for $x\ge1$, while at small
$x$ they lie above, by an amount that grows as $x$ and $\bar R$ decrease: up to $8.6\,\%$ for $\bar R=1000$ and $16\,\%$ for $\bar R=116$ in
the range shown. The statistical errors are below $0.35\,\%$.

\begin{figure*}[htb]
\includegraphics[width=\textwidth]{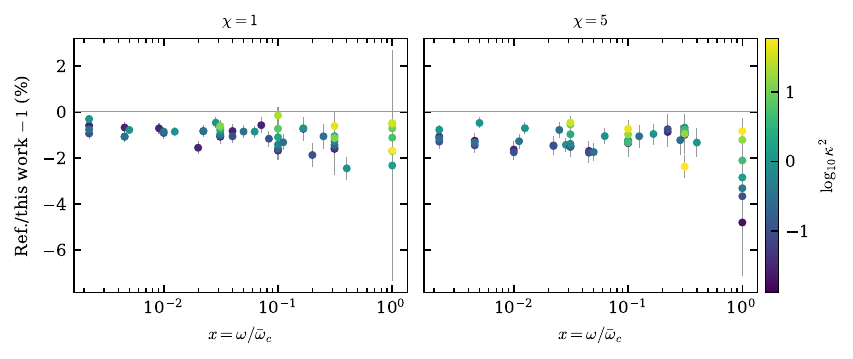}
\caption{Spectrum differential in $\kappa^2$: relative difference between the curves of Figs.~1 ($\chi=1$) and~2 ($\chi=5$) of
Ref.~\cite{AndresApolinarioDominguez2020} and this work, Ref.~\cite{AndresApolinarioDominguez2020}/this work $-\,1$ in \%, at the published points; the color gives $\kappa^2$. The figure was produced with scripts written with the assistance of an AI tool (see Sec.~\ref{sec:validation} and Acknowledgments).}\label{fig:differential}
\end{figure*}

\textit{Differential spectrum.} Figure~\ref{fig:differential} compares the spectrum differential in $\kappa^2$ with the curves of
Figs.~1 and~2 of Ref.~\cite{AndresApolinarioDominguez2020}, for $\chi=1$ and $\chi=5$, at 56 points each. Their values lie below ours by $1.05\,\%$ on average
(rms $1.1\,\%$) for $\chi=1$ and by $1.4\,\%$ (rms $1.6\,\%$, maximum $4.8\,\%$) for $\chi=5$, with a mean statistical error of $0.5\,\%$.

\begin{figure*}[htb]
\includegraphics[width=\textwidth]{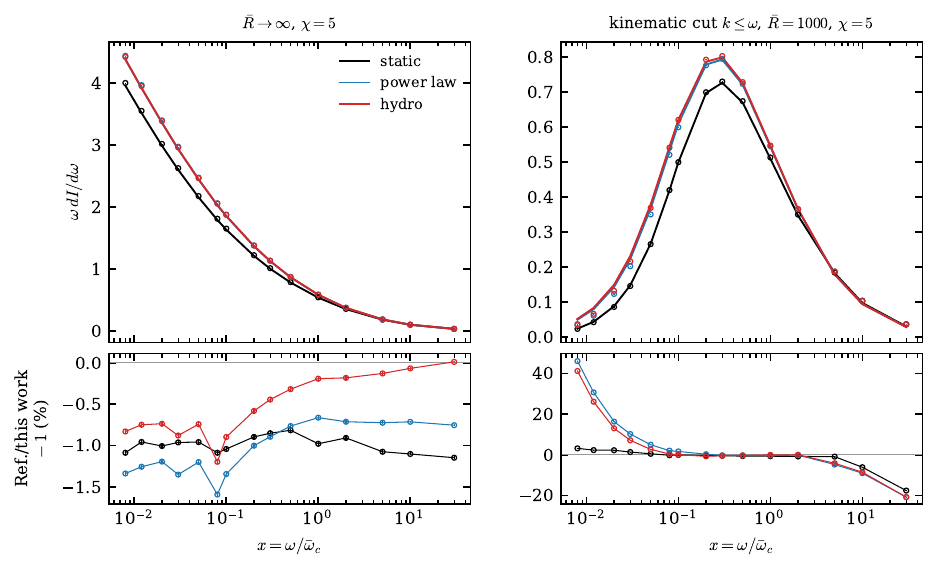}
\caption{Static, power-law and hydrodynamic media with $\chi=5$, without cut (left) and with the kinematic cut $k\le\omega$ and
$\bar R=1000$ (right). Lines: Fig.~3 of Ref.~\cite{Andres2024}; points: this work. Lower panels: relative difference, Ref.~\cite{Andres2024}/this work $-\,1$, in \%. The figure was produced with scripts written with the assistance of an AI tool (see Sec.~\ref{sec:validation} and Acknowledgments).}\label{fig:expanding}
\end{figure*}

\textit{Expanding media.} Figure~\ref{fig:expanding} compares our results with Fig.~3 of Ref.~\cite{Andres2024} for $\chi=5$: a static medium, a
power-law profile, and the hydrodynamic profile read from the inset of that figure, with time measured from $\tau_{\rm hydro}$; with this
choice the equivalent static medium has $\bar R=1029$, close to the value $\bar R=1000$ quoted there. Following that reference, the power-law profile is $n(t)=\bar n_0/(t+\bar t_0)^\alpha$, $\mu^2(t)=\bar\mu_0^2/(t+\bar t_0)^{2\alpha}$ with $\alpha=0.5$ and $\bar t_0/\bar L=0.1$, normalized to $\chi=5$, and the hydrodynamic one is $n=k_1T$, $\mu^2=k_2T^2$ with $k_1=0.5$ and $k_2=45$ along the temperature profile $T(t)$ of the inset, which gives $\chi=4.97$. Without cut, their curves lie
$0$--$1.6\,\%$ below our results. With the kinematic cut, at small $x$ their curves for the expanding media lie well above ours:
by $46\,\%$ (power law) and $41\,\%$ (hydrodynamic profile) at $x=0.008$, decreasing to $16\,\%$ and $13\,\%$ at $x=0.02$ and below $2\,\%$ for
$x\ge0.1$; for the static medium the difference at $x=0.008$ is $3\,\%$. At large $x$ their curves with cut fall below ours, by
$4$--$9\,\%$ at $x=5$--$10$ and by about $20\,\%$ at $x=30$ for the expanding media ($6\,\%$ and $18\,\%$ at $x=10$ and $30$ for the static
one). The statistical errors are below $0.25\,\%$.

\section{Summary and outlook}\label{sec:summary}
We have computed the BDMPS-Z spectrum in the soft-gluon limit for a Yukawa potential, to all orders in opacity, by writing the
Yukawa dipole cross section as a Laplace superposition of Gaussians. This turns the medium into a Poisson gas of Gaussian scattering
centers, each configuration of which is a harmonic oscillator with an impulsive frequency, solved in closed form. The result is a master
formula for the spectrum differential in $k_\perp$ and with a cut on $k_\perp$, valid for arbitrary profiles $n(s)$ and $\mu(s)$, whose
$K\to\infty$ limit is the average of the logarithm of a Gelfand--Yaglom determinant. Its Monte Carlo evaluation is efficient, gives the
whole $k_\perp$ dependence from a single run, and is trivially parallelizable. It reproduces the first order in opacity and the published
results without cut, while with a kinematic cut we find sizeable relative deviations from the published results, both at small gluon energy fraction,
where the cut strongly suppresses the spectrum, and at large energy fraction, where they reach 5--20\,\%.

We expect the construction not to rely on the soft limit in an essential way: what it requires is that, configuration by
configuration, the problem is Gaussian. We expect it to extend to the more complex building blocks of an in-medium parton cascade, such as finite energy
fraction and emissions off color antennas (for the latter, see Ref.~\cite{Andres2026antenna} for a recent numerical treatment). The master formula also provides a natural starting point for analytical approximations,
which we will address in future work. Finally, the independence of the samples makes the method well suited to massively parallel
hardware.

\section*{Data availability}
The code used for all numerical results of this article is openly available at \url{https://github.com/manoel-maker/poisson-gas-bdmps_z}.
The scripts that produce the figures and the curves read from Refs.~\cite{AndresApolinarioDominguez2020,Andres2024} are available
from the author upon reasonable request.

\begin{acknowledgments}
The author used Claude (Anthropic; several model versions, through claude.ai and Claude Code, between June and November 2026)
as a research assistant. The representation of the Yukawa dipole cross section as a Laplace superposition of Gaussians, the central
derivations, the choice of what to compute and to include, and the conclusions are the author's; the Monte Carlo strategy was developed
in the work with Claude. Claude was used to check derivations, including
independent numerical checks; to write and validate the Monte Carlo code and the scripts that produce the figures (see
Sec.~\ref{sec:validation}); to search for and verify references; and to draft the English text of parts of the manuscript from the
author's notes. The abstract and Secs.~\ref{sec:intro} and~\ref{sec:representation} follow the author's own drafts. The author checked the derivations by hand, reviewed the code and the text, and takes full
responsibility for the content.
\end{acknowledgments}

\appendix
\section{Continuous route: Jacobi equation and Arnold's propagator}\label{app:arnold}
The exponent of $\K_c$ in Eq.~\eqref{eq:Kconf} is $\int[\frac{i\omega}{2}\dot\vrr^{\,2}-U(s)r^2]$ with $U=\sum_j\tau_j\delta(s-s_j)$.
The classical equation is $\ddot\vrr=\sum_ja_j\delta(s-s_j)\,\vrr$, with $a_j$ as in Eq.~\eqref{eq:aj}; in Arnold's notation \cite{Arnold2009},
$\ddot c=-\omega_0^2(t)c$ with $\omega_0^2(t)=-\sum_ja_j\delta(t-s_j)$. The Jacobi solution $S(t_2;t_1)$ satisfies
\begin{equation}\label{eq:jacobi}
\partial_{t_2}^2S=\sum_ja_j\,\delta(t_2-s_j)\,S,
\end{equation}
with $S(t_1;t_1)=0$ and $\partial_{t_2}S|_{t_2=t_1}=1$.
Between two centers $S$ is linear in $t_2$, and at each center $s_j>t_1$ it is continuous while its slope jumps as in
Eq.~\eqref{eq:salto}. The products of $\delta$ with the solution are well defined without any prescription: $S$ is continuous at
$s_j$, the discontinuity is in $\Sd$, which never multiplies a $\delta$, and integrating Eq.~\eqref{eq:jacobi} over
$[s_j-\epsilon,s_j+\epsilon]$ gives Eq.~\eqref{eq:salto}.

Since the exponent is quadratic, the path integral is the exponential of the classical action times the Van Vleck prefactor. In two
transverse dimensions,
\begin{multline}\label{eq:prop}
\K_c(t_2,\vz;t_1,\vy)=\frac{\omega}{2\pi i\,S}\\
\times\exp\Big\{\frac{i\omega}{2S}\big[\Sd\,z^2+C\,y^2-2\,\vz\cdot\vy\big]\Big\},
\end{multline}
with $S=S(t_2;t_1)$ and $C=-\partial_{t_1}S$. For $\vy=0$ the classical path is $\vrr(s)=\vz\,S(s;t_1)/S(t_2;t_1)$ and the classical
action is $\frac{\omega}{2}z^2\Sd/S$. Comparing with Eq.~\eqref{eq:kacum} at $t=t_2$ gives $q=1/S$, $p=\Sd/S$ and
$\mathcal N=\omega/(2\pi iS)$, which is Eq.~\eqref{eq:pqS}: both routes define the same $S$. Then
$\nabla_{\vy}\K_c|_{\vy=0}=-\frac{\omega^2}{2\pi S^2}\vz\,e^{\eta z^2}$, which is Eq.~(3.3) of Arnold \cite{Arnold2009} with $M=\omega$,
$\mathbf B=\vz$, and Eq.~\eqref{eq:dd} follows. In this route the continuity of $\vartheta$ is checked directly: when $t_2$ crosses
$s_j$, $T$ decreases by $\tau_j$ and $\Sd$ increases by $a_jS_j$, so that $\eta$ changes by $\frac{i\omega}{2}a_j=-\tau_j$ and
$\vartheta=T-\eta$ does not change.

\section{HTL potential with a magnetic mass}\label{app:htl}
The HTL potential~\cite{AurencheGelisZaraket2002} is, up to a constant prefactor $C$,
\begin{equation}
V(q)=C\,\frac{m_D^2}{q^2(q^2+m_D^2)}=C\Big[\frac1{q^2}-\frac1{q^2+m_D^2}\Big].
\end{equation}
The term $1/q^2$, from the unscreened magnetostatic gluons, makes the total elastic cross section logarithmically divergent in the
infrared, so that the Poisson intensity of Sec.~\ref{sec:poisson} would be infinite. A magnetic mass $m_M$ regularizes it,
\begin{align}\label{eq:htlM}
V_M(q)&=C\,\frac{m_D^2-m_M^2}{(q^2+m_M^2)(q^2+m_D^2)}\notag\\
&=C\Big[\frac1{q^2+m_M^2}-\frac1{q^2+m_D^2}\Big],
\end{align}
which has the form of the effective cross section of Ref.~\cite{DjordjevicDjordjevic2012} (their Eq.~(9), with $\mu_M=m_M$, $\mu_E=m_D$).
Writing $1/(q^2+m^2)=\int_0^\infty\dd\lambda\,e^{-\lambda(q^2+m^2)}$,
\begin{equation}
V_M(q)=C\int_0^\infty\!\dd\lambda\;e^{-\lambda q^2}\big[e^{-\lambda m_M^2}-e^{-\lambda m_D^2}\big],
\end{equation}
a superposition of Gaussians whose weight is positive for every $\lambda$ if and only if $m_M<m_D$. Performing the Gaussian integral
over $\mathbf q$ and changing variables to $\tau=1/(4\lambda)$,
\begin{align}
\sigma(r)&=\frac{C}{4\pi}\int_0^\infty\!\frac{\dd\tau}{\tau}\big[e^{-m_M^2/4\tau}-e^{-m_D^2/4\tau}\big]\big(1-e^{-\tau r^2}\big)\notag\\
&=\frac{C}{2\pi}\Big[\ln\frac{m_D}{m_M}-K_0(m_Mr)+K_0(m_Dr)\Big].
\end{align}
The total cross section $\sigma(\infty)=\frac{C}{4\pi}\ln(m_D^2/m_M^2)$ is finite, and $v=\frac12n\sigma=n_M\int\dd\tau\,\rho_M(\tau)\,(1-e^{-\tau r^2})$ with
$n_M=\frac12n\,\sigma(\infty)$ and
\begin{equation}
\rho_M(\tau)=\frac{e^{-m_M^2/4\tau}-e^{-m_D^2/4\tau}}{\tau\,\ln(m_D^2/m_M^2)} .
\end{equation}
This is a probability density, positive if and only if $m_M<m_D$, with the same Coulomb tail as in the Yukawa case,
$\rho_M\simeq(m_D^2-m_M^2)/[4\tau^2\ln(m_D^2/m_M^2)]$ for $\tau\to\infty$. Everything in Secs.~\ref{sec:poisson}--\ref{sec:mc} then applies with
$n\to n_M$ and $\rho_s\to\rho_M$: only the sampling of the widths changes. The condition $\mu_M<\mu_E$ imposed in
Ref.~\cite{DjordjevicDjordjevic2012} is precisely the positivity of this Poisson gas. For $m_M\to0$ the intensity diverges
logarithmically: the unregularized HTL potential is still a superposition of Gaussians with positive weights, but not a Poisson gas
with finite intensity. The magnetic mass is of order $g^2T$ and nonperturbative; Ref.~\cite{DjordjevicDjordjevic2012} takes
$0.4<\mu_M/\mu_E<0.6$ from nonperturbative estimates.

\bibliography{calvo_poisson_gas}

\begin{thebibliography}{27}%
\makeatletter
\providecommand \@ifxundefined [1]{%
 \@ifx{#1\undefined}
}%
\providecommand \@ifnum [1]{%
 \ifnum #1\expandafter \@firstoftwo
 \else \expandafter \@secondoftwo
 \fi
}%
\providecommand \@ifx [1]{%
 \ifx #1\expandafter \@firstoftwo
 \else \expandafter \@secondoftwo
 \fi
}%
\providecommand \natexlab [1]{#1}%
\providecommand \enquote  [1]{``#1''}%
\providecommand \bibnamefont  [1]{#1}%
\providecommand \bibfnamefont [1]{#1}%
\providecommand \citenamefont [1]{#1}%
\providecommand \href@noop [0]{\@secondoftwo}%
\providecommand \href [0]{\begingroup \@sanitize@url \@href}%
\providecommand \@href[1]{\@@startlink{#1}\@@href}%
\providecommand \@@href[1]{\endgroup#1\@@endlink}%
\providecommand \@sanitize@url [0]{\catcode `\\12\catcode `\$12\catcode
  `\&12\catcode `\#12\catcode `\^12\catcode `\_12\catcode `\%12\relax}%
\providecommand \@@startlink[1]{}%
\providecommand \@@endlink[0]{}%
\providecommand \url  [0]{\begingroup\@sanitize@url \@url }%
\providecommand \@url [1]{\endgroup\@href {#1}{\urlprefix }}%
\providecommand \urlprefix  [0]{URL }%
\providecommand \Eprint [0]{\href }%
\providecommand \doibase [0]{https://doi.org/}%
\providecommand \selectlanguage [0]{\@gobble}%
\providecommand \bibinfo  [0]{\@secondoftwo}%
\providecommand \bibfield  [0]{\@secondoftwo}%
\providecommand \translation [1]{[#1]}%
\providecommand \BibitemOpen [0]{}%
\providecommand \bibitemStop [0]{}%
\providecommand \bibitemNoStop [0]{.\EOS\space}%
\providecommand \EOS [0]{\spacefactor3000\relax}%
\providecommand \BibitemShut  [1]{\csname bibitem#1\endcsname}%
\let\auto@bib@innerbib\@empty
%</preamble>
\bibitem [{\citenamefont {Baier}\ \emph
  {et~al.}(1997{\natexlab{a}})\citenamefont {Baier}, \citenamefont
  {Dokshitzer}, \citenamefont {Mueller}, \citenamefont {Peign\'e},\ and\
  \citenamefont {Schiff}}]{BDMPS1997a}%
  \BibitemOpen
  \bibfield  {author} {\bibinfo {author} {\bibfnamefont {R.}~\bibnamefont
  {Baier}}, \bibinfo {author} {\bibfnamefont {Y.~L.}\ \bibnamefont
  {Dokshitzer}}, \bibinfo {author} {\bibfnamefont {A.~H.}\ \bibnamefont
  {Mueller}}, \bibinfo {author} {\bibfnamefont {S.}~\bibnamefont {Peign\'e}},\
  and\ \bibinfo {author} {\bibfnamefont {D.}~\bibnamefont {Schiff}},\
  }\bibfield  {title} {\bibinfo {title} {Radiative energy loss of high-energy
  quarks and gluons in a finite volume quark-gluon plasma},\ }\href@noop {}
  {\bibfield  {journal} {\bibinfo  {journal} {Nucl. Phys. B}\ }\textbf
  {\bibinfo {volume} {483}},\ \bibinfo {pages} {291} (\bibinfo {year}
  {1997}{\natexlab{a}})},\ \Eprint {https://arxiv.org/abs/hep-ph/9607355}
  {hep-ph/9607355} \BibitemShut {NoStop}%
\bibitem [{\citenamefont {Baier}\ \emph
  {et~al.}(1997{\natexlab{b}})\citenamefont {Baier}, \citenamefont
  {Dokshitzer}, \citenamefont {Mueller}, \citenamefont {Peign\'e},\ and\
  \citenamefont {Schiff}}]{BDMPS1997b}%
  \BibitemOpen
  \bibfield  {author} {\bibinfo {author} {\bibfnamefont {R.}~\bibnamefont
  {Baier}}, \bibinfo {author} {\bibfnamefont {Y.~L.}\ \bibnamefont
  {Dokshitzer}}, \bibinfo {author} {\bibfnamefont {A.~H.}\ \bibnamefont
  {Mueller}}, \bibinfo {author} {\bibfnamefont {S.}~\bibnamefont {Peign\'e}},\
  and\ \bibinfo {author} {\bibfnamefont {D.}~\bibnamefont {Schiff}},\
  }\bibfield  {title} {\bibinfo {title} {Radiative energy loss and $p_t$
  broadening of high-energy partons in nuclei},\ }\href@noop {} {\bibfield
  {journal} {\bibinfo  {journal} {Nucl. Phys. B}\ }\textbf {\bibinfo {volume}
  {484}},\ \bibinfo {pages} {265} (\bibinfo {year} {1997}{\natexlab{b}})},\
  \Eprint {https://arxiv.org/abs/hep-ph/9608322} {hep-ph/9608322} \BibitemShut
  {NoStop}%
\bibitem [{\citenamefont {Zakharov}(1996)}]{Zakharov1996}%
  \BibitemOpen
  \bibfield  {author} {\bibinfo {author} {\bibfnamefont {B.~G.}\ \bibnamefont
  {Zakharov}},\ }\bibfield  {title} {\bibinfo {title} {Fully quantum treatment
  of the {Landau-Pomeranchuk-Migdal} effect in {QED} and {QCD}},\ }\href@noop
  {} {\bibfield  {journal} {\bibinfo  {journal} {JETP Lett.}\ }\textbf
  {\bibinfo {volume} {63}},\ \bibinfo {pages} {952} (\bibinfo {year} {1996})},\
  \Eprint {https://arxiv.org/abs/hep-ph/9607440} {hep-ph/9607440} \BibitemShut
  {NoStop}%
\bibitem [{\citenamefont {Zakharov}(1997)}]{Zakharov1997}%
  \BibitemOpen
  \bibfield  {author} {\bibinfo {author} {\bibfnamefont {B.~G.}\ \bibnamefont
  {Zakharov}},\ }\bibfield  {title} {\bibinfo {title} {Radiative energy loss of
  high-energy quarks in finite size nuclear matter and quark gluon plasma},\
  }\href@noop {} {\bibfield  {journal} {\bibinfo  {journal} {JETP Lett.}\
  }\textbf {\bibinfo {volume} {65}},\ \bibinfo {pages} {615} (\bibinfo {year}
  {1997})},\ \Eprint {https://arxiv.org/abs/hep-ph/9704255} {hep-ph/9704255}
  \BibitemShut {NoStop}%
\bibitem [{\citenamefont {Andres}\ \emph {et~al.}(2020)\citenamefont {Andres},
  \citenamefont {Apolin\'ario},\ and\ \citenamefont
  {Dominguez}}]{AndresApolinarioDominguez2020}%
  \BibitemOpen
  \bibfield  {author} {\bibinfo {author} {\bibfnamefont {C.}~\bibnamefont
  {Andres}}, \bibinfo {author} {\bibfnamefont {L.}~\bibnamefont
  {Apolin\'ario}},\ and\ \bibinfo {author} {\bibfnamefont {F.}~\bibnamefont
  {Dominguez}},\ }\bibfield  {title} {\bibinfo {title} {Medium-induced gluon
  radiation with full resummation of multiple scatterings for realistic
  parton-medium interactions},\ }\href@noop {} {\bibfield  {journal} {\bibinfo
  {journal} {JHEP}\ }\textbf {\bibinfo {volume} {07 (2020)}},\ \bibinfo {pages}
  {114}},\ \Eprint {https://arxiv.org/abs/2002.01517} {2002.01517} \BibitemShut
  {NoStop}%
\bibitem [{\citenamefont {Gyulassy}\ and\ \citenamefont
  {Wang}(1994)}]{GyulassyWang1994}%
  \BibitemOpen
  \bibfield  {author} {\bibinfo {author} {\bibfnamefont {M.}~\bibnamefont
  {Gyulassy}}\ and\ \bibinfo {author} {\bibfnamefont {X.-N.}\ \bibnamefont
  {Wang}},\ }\bibfield  {title} {\bibinfo {title} {Multiple collisions and
  induced gluon bremsstrahlung in {QCD}},\ }\href@noop {} {\bibfield  {journal}
  {\bibinfo  {journal} {Nucl. Phys. B}\ }\textbf {\bibinfo {volume} {420}},\
  \bibinfo {pages} {583} (\bibinfo {year} {1994})},\ \Eprint
  {https://arxiv.org/abs/nucl-th/9306003} {nucl-th/9306003} \BibitemShut
  {NoStop}%
\bibitem [{\citenamefont {Aurenche}\ \emph {et~al.}(2002)\citenamefont
  {Aurenche}, \citenamefont {Gelis},\ and\ \citenamefont
  {Zaraket}}]{AurencheGelisZaraket2002}%
  \BibitemOpen
  \bibfield  {author} {\bibinfo {author} {\bibfnamefont {P.}~\bibnamefont
  {Aurenche}}, \bibinfo {author} {\bibfnamefont {F.}~\bibnamefont {Gelis}},\
  and\ \bibinfo {author} {\bibfnamefont {H.}~\bibnamefont {Zaraket}},\
  }\bibfield  {title} {\bibinfo {title} {A simple sum rule for the thermal
  gluon spectral function and applications},\ }\href@noop {} {\bibfield
  {journal} {\bibinfo  {journal} {JHEP}\ }\textbf {\bibinfo {volume} {05
  (2002)}},\ \bibinfo {pages} {043}},\ \Eprint
  {https://arxiv.org/abs/hep-ph/0204146} {hep-ph/0204146} \BibitemShut
  {NoStop}%
\bibitem [{\citenamefont {Gyulassy}\ \emph {et~al.}(2001)\citenamefont
  {Gyulassy}, \citenamefont {Levai},\ and\ \citenamefont {Vitev}}]{GLV2001}%
  \BibitemOpen
  \bibfield  {author} {\bibinfo {author} {\bibfnamefont {M.}~\bibnamefont
  {Gyulassy}}, \bibinfo {author} {\bibfnamefont {P.}~\bibnamefont {Levai}},\
  and\ \bibinfo {author} {\bibfnamefont {I.}~\bibnamefont {Vitev}},\ }\bibfield
   {title} {\bibinfo {title} {Reaction operator approach to non-{Abelian}
  energy loss},\ }\href@noop {} {\bibfield  {journal} {\bibinfo  {journal}
  {Nucl. Phys. B}\ }\textbf {\bibinfo {volume} {594}},\ \bibinfo {pages} {371}
  (\bibinfo {year} {2001})},\ \Eprint {https://arxiv.org/abs/nucl-th/0006010}
  {nucl-th/0006010} \BibitemShut {NoStop}%
\bibitem [{\citenamefont {Gyulassy}\ \emph {et~al.}(2000)\citenamefont
  {Gyulassy}, \citenamefont {Levai},\ and\ \citenamefont {Vitev}}]{GLV2000}%
  \BibitemOpen
  \bibfield  {author} {\bibinfo {author} {\bibfnamefont {M.}~\bibnamefont
  {Gyulassy}}, \bibinfo {author} {\bibfnamefont {P.}~\bibnamefont {Levai}},\
  and\ \bibinfo {author} {\bibfnamefont {I.}~\bibnamefont {Vitev}},\ }\bibfield
   {title} {\bibinfo {title} {Non-{Abelian} energy loss at finite opacity},\
  }\href@noop {} {\bibfield  {journal} {\bibinfo  {journal} {Phys. Rev. Lett.}\
  }\textbf {\bibinfo {volume} {85}},\ \bibinfo {pages} {5535} (\bibinfo {year}
  {2000})},\ \Eprint {https://arxiv.org/abs/nucl-th/0005032} {nucl-th/0005032}
  \BibitemShut {NoStop}%
\bibitem [{\citenamefont {Wiedemann}(2000)}]{Wiedemann2000}%
  \BibitemOpen
  \bibfield  {author} {\bibinfo {author} {\bibfnamefont {U.~A.}\ \bibnamefont
  {Wiedemann}},\ }\bibfield  {title} {\bibinfo {title} {Gluon radiation off
  hard quarks in a nuclear environment: opacity expansion},\ }\href@noop {}
  {\bibfield  {journal} {\bibinfo  {journal} {Nucl. Phys. B}\ }\textbf
  {\bibinfo {volume} {588}},\ \bibinfo {pages} {303} (\bibinfo {year}
  {2000})},\ \Eprint {https://arxiv.org/abs/hep-ph/0005129} {hep-ph/0005129}
  \BibitemShut {NoStop}%
\bibitem [{\citenamefont {Salgado}\ and\ \citenamefont
  {Wiedemann}(2003)}]{SalgadoWiedemann2003}%
  \BibitemOpen
  \bibfield  {author} {\bibinfo {author} {\bibfnamefont {C.~A.}\ \bibnamefont
  {Salgado}}\ and\ \bibinfo {author} {\bibfnamefont {U.~A.}\ \bibnamefont
  {Wiedemann}},\ }\bibfield  {title} {\bibinfo {title} {Calculating quenching
  weights},\ }\href@noop {} {\bibfield  {journal} {\bibinfo  {journal} {Phys.
  Rev. D}\ }\textbf {\bibinfo {volume} {68}},\ \bibinfo {pages} {014008}
  (\bibinfo {year} {2003})},\ \Eprint {https://arxiv.org/abs/hep-ph/0302184}
  {hep-ph/0302184} \BibitemShut {NoStop}%
\bibitem [{\citenamefont {Mehtar-Tani}(2019)}]{MehtarTani2019}%
  \BibitemOpen
  \bibfield  {author} {\bibinfo {author} {\bibfnamefont {Y.}~\bibnamefont
  {Mehtar-Tani}},\ }\bibfield  {title} {\bibinfo {title} {Gluon bremsstrahlung
  in finite media beyond multiple soft scattering approximation},\ }\href@noop
  {} {\bibfield  {journal} {\bibinfo  {journal} {JHEP}\ }\textbf {\bibinfo
  {volume} {07 (2019)}},\ \bibinfo {pages} {057}},\ \Eprint
  {https://arxiv.org/abs/1903.00506} {1903.00506} \BibitemShut {NoStop}%
\bibitem [{\citenamefont {Mehtar-Tani}\ and\ \citenamefont
  {Tywoniuk}(2020)}]{MehtarTaniTywoniuk2020}%
  \BibitemOpen
  \bibfield  {author} {\bibinfo {author} {\bibfnamefont {Y.}~\bibnamefont
  {Mehtar-Tani}}\ and\ \bibinfo {author} {\bibfnamefont {K.}~\bibnamefont
  {Tywoniuk}},\ }\bibfield  {title} {\bibinfo {title} {Improved opacity
  expansion for medium-induced parton splitting},\ }\href@noop {} {\bibfield
  {journal} {\bibinfo  {journal} {JHEP}\ }\textbf {\bibinfo {volume} {06
  (2020)}},\ \bibinfo {pages} {187}},\ \Eprint
  {https://arxiv.org/abs/1910.02032} {1910.02032} \BibitemShut {NoStop}%
\bibitem [{\citenamefont {Barata}\ and\ \citenamefont
  {Mehtar-Tani}(2020)}]{BarataMehtarTani2020}%
  \BibitemOpen
  \bibfield  {author} {\bibinfo {author} {\bibfnamefont {J.}~\bibnamefont
  {Barata}}\ and\ \bibinfo {author} {\bibfnamefont {Y.}~\bibnamefont
  {Mehtar-Tani}},\ }\bibfield  {title} {\bibinfo {title} {Improved opacity
  expansion at {NNLO} for medium induced gluon radiation},\ }\href@noop {}
  {\bibfield  {journal} {\bibinfo  {journal} {JHEP}\ }\textbf {\bibinfo
  {volume} {10 (2020)}},\ \bibinfo {pages} {176}},\ \Eprint
  {https://arxiv.org/abs/2004.02323} {2004.02323} \BibitemShut {NoStop}%
\bibitem [{\citenamefont {Isaksen}\ \emph {et~al.}(2023)\citenamefont
  {Isaksen}, \citenamefont {Takacs},\ and\ \citenamefont
  {Tywoniuk}}]{IsaksenTakacsTywoniuk2023}%
  \BibitemOpen
  \bibfield  {author} {\bibinfo {author} {\bibfnamefont {J.~H.}\ \bibnamefont
  {Isaksen}}, \bibinfo {author} {\bibfnamefont {A.}~\bibnamefont {Takacs}},\
  and\ \bibinfo {author} {\bibfnamefont {K.}~\bibnamefont {Tywoniuk}},\
  }\bibfield  {title} {\bibinfo {title} {A unified picture of medium-induced
  radiation},\ }\href@noop {} {\bibfield  {journal} {\bibinfo  {journal}
  {JHEP}\ }\textbf {\bibinfo {volume} {02 (2023)}},\ \bibinfo {pages} {156}},\
  \Eprint {https://arxiv.org/abs/2206.02811} {2206.02811} \BibitemShut
  {NoStop}%
\bibitem [{\citenamefont {Caron-Huot}\ and\ \citenamefont
  {Gale}(2010)}]{CaronHuotGale2010}%
  \BibitemOpen
  \bibfield  {author} {\bibinfo {author} {\bibfnamefont {S.}~\bibnamefont
  {Caron-Huot}}\ and\ \bibinfo {author} {\bibfnamefont {C.}~\bibnamefont
  {Gale}},\ }\bibfield  {title} {\bibinfo {title} {Finite-size effects on the
  radiative energy loss of a fast parton in hot and dense strongly interacting
  matter},\ }\href@noop {} {\bibfield  {journal} {\bibinfo  {journal} {Phys.
  Rev. C}\ }\textbf {\bibinfo {volume} {82}},\ \bibinfo {pages} {064902}
  (\bibinfo {year} {2010})},\ \Eprint {https://arxiv.org/abs/1006.2379}
  {1006.2379} \BibitemShut {NoStop}%
\bibitem [{\citenamefont {Andres}\ \emph {et~al.}(2024)\citenamefont {Andres},
  \citenamefont {Apolin\'ario}, \citenamefont {Dominguez},\ and\ \citenamefont
  {Gonzalez~Martinez}}]{Andres2024}%
  \BibitemOpen
  \bibfield  {author} {\bibinfo {author} {\bibfnamefont {C.}~\bibnamefont
  {Andres}}, \bibinfo {author} {\bibfnamefont {L.}~\bibnamefont
  {Apolin\'ario}}, \bibinfo {author} {\bibfnamefont {F.}~\bibnamefont
  {Dominguez}},\ and\ \bibinfo {author} {\bibfnamefont {M.}~\bibnamefont
  {Gonzalez~Martinez}},\ }\bibfield  {title} {\bibinfo {title} {In-medium gluon
  radiation spectrum with all-order resummation of multiple scatterings in
  longitudinally evolving media},\ }\href@noop {} {\bibfield  {journal}
  {\bibinfo  {journal} {JHEP}\ }\textbf {\bibinfo {volume} {11 (2024)}},\
  \bibinfo {pages} {025}},\ \Eprint {https://arxiv.org/abs/2307.06226}
  {2307.06226} \BibitemShut {NoStop}%
\bibitem [{\citenamefont {Feal}\ and\ \citenamefont
  {Vazquez}(2018)}]{FealVazquez2018}%
  \BibitemOpen
  \bibfield  {author} {\bibinfo {author} {\bibfnamefont {X.}~\bibnamefont
  {Feal}}\ and\ \bibinfo {author} {\bibfnamefont {R.}~\bibnamefont {Vazquez}},\
  }\bibfield  {title} {\bibinfo {title} {Intensity of gluon bremsstrahlung in a
  finite plasma},\ }\href@noop {} {\bibfield  {journal} {\bibinfo  {journal}
  {Phys. Rev. D}\ }\textbf {\bibinfo {volume} {98}},\ \bibinfo {pages} {074029}
  (\bibinfo {year} {2018})},\ \Eprint {https://arxiv.org/abs/1811.01591}
  {1811.01591} \BibitemShut {NoStop}%
\bibitem [{\citenamefont {Arnold}(2009)}]{Arnold2009}%
  \BibitemOpen
  \bibfield  {author} {\bibinfo {author} {\bibfnamefont {P.~B.}\ \bibnamefont
  {Arnold}},\ }\bibfield  {title} {\bibinfo {title} {Simple formula for
  high-energy gluon bremsstrahlung in a finite, expanding medium},\ }\href@noop
  {} {\bibfield  {journal} {\bibinfo  {journal} {Phys. Rev. D}\ }\textbf
  {\bibinfo {volume} {79}},\ \bibinfo {pages} {065025} (\bibinfo {year}
  {2009})},\ \Eprint {https://arxiv.org/abs/0808.2767} {0808.2767} \BibitemShut
  {NoStop}%
\bibitem [{\citenamefont {Widder}(1941)}]{Widder1941}%
  \BibitemOpen
  \bibfield  {author} {\bibinfo {author} {\bibfnamefont {D.~V.}\ \bibnamefont
  {Widder}},\ }\href@noop {} {\emph {\bibinfo {title} {The {Laplace}
  Transform}}},\ \bibinfo {series} {Princeton Mathematical Series},
  Vol.~\bibinfo {volume} {6}\ (\bibinfo  {publisher} {Princeton University
  Press},\ \bibinfo {address} {Princeton},\ \bibinfo {year} {1941})\BibitemShut
  {NoStop}%
\bibitem [{\citenamefont {Schwinger}(1951)}]{Schwinger1951}%
  \BibitemOpen
  \bibfield  {author} {\bibinfo {author} {\bibfnamefont {J.}~\bibnamefont
  {Schwinger}},\ }\bibfield  {title} {\bibinfo {title} {On gauge invariance and
  vacuum polarization},\ }\href {https://doi.org/10.1103/PhysRev.82.664}
  {\bibfield  {journal} {\bibinfo  {journal} {Phys. Rev.}\ }\textbf {\bibinfo
  {volume} {82}},\ \bibinfo {pages} {664} (\bibinfo {year} {1951})}\BibitemShut
  {NoStop}%
\bibitem [{\citenamefont {Last}\ and\ \citenamefont
  {Penrose}(2017)}]{LastPenrose2017}%
  \BibitemOpen
  \bibfield  {author} {\bibinfo {author} {\bibfnamefont {G.}~\bibnamefont
  {Last}}\ and\ \bibinfo {author} {\bibfnamefont {M.}~\bibnamefont {Penrose}},\
  }\href@noop {} {\emph {\bibinfo {title} {Lectures on the {Poisson}
  Process}}},\ IMS Textbooks\ (\bibinfo  {publisher} {Cambridge University
  Press},\ \bibinfo {address} {Cambridge},\ \bibinfo {year} {2017})\BibitemShut
  {NoStop}%
\bibitem [{\citenamefont {Gel'fand}\ and\ \citenamefont
  {Yaglom}(1960)}]{GelfandYaglom1960}%
  \BibitemOpen
  \bibfield  {author} {\bibinfo {author} {\bibfnamefont {I.~M.}\ \bibnamefont
  {Gel'fand}}\ and\ \bibinfo {author} {\bibfnamefont {A.~M.}\ \bibnamefont
  {Yaglom}},\ }\bibfield  {title} {\bibinfo {title} {Integration in functional
  spaces and its applications in quantum physics},\ }\href@noop {} {\bibfield
  {journal} {\bibinfo  {journal} {J. Math. Phys.}\ }\textbf {\bibinfo {volume}
  {1}},\ \bibinfo {pages} {48} (\bibinfo {year} {1960})}\BibitemShut {NoStop}%
\bibitem [{\citenamefont {James}(1980)}]{James1980}%
  \BibitemOpen
  \bibfield  {author} {\bibinfo {author} {\bibfnamefont {F.}~\bibnamefont
  {James}},\ }\bibfield  {title} {\bibinfo {title} {{Monte Carlo} theory and
  practice},\ }\href {https://doi.org/10.1088/0034-4885/43/9/002} {\bibfield
  {journal} {\bibinfo  {journal} {Rep. Prog. Phys.}\ }\textbf {\bibinfo
  {volume} {43}},\ \bibinfo {pages} {1145} (\bibinfo {year}
  {1980})}\BibitemShut {NoStop}%
\bibitem [{\citenamefont {Mecke}(1967)}]{Mecke1967}%
  \BibitemOpen
  \bibfield  {author} {\bibinfo {author} {\bibfnamefont {J.}~\bibnamefont
  {Mecke}},\ }\bibfield  {title} {\bibinfo {title} {Station\"are zuf\"allige
  {Ma\ss}e auf lokalkompakten {Abelschen} {Gruppen}},\ }\href
  {https://doi.org/10.1007/BF00535466} {\bibfield  {journal} {\bibinfo
  {journal} {Z. Wahrsch. verw. Geb.}\ }\textbf {\bibinfo {volume} {9}},\
  \bibinfo {pages} {36} (\bibinfo {year} {1967})}\BibitemShut {NoStop}%
\bibitem [{\citenamefont {Andres}\ \emph {et~al.}(2026)\citenamefont {Andres},
  \citenamefont {Apolin\'ario}, \citenamefont {Armesto}, \citenamefont
  {Cordeiro}, \citenamefont {Dom\'inguez}, \citenamefont
  {Guerrero-Rodr\'iguez},\ and\ \citenamefont {Milhano}}]{Andres2026antenna}%
  \BibitemOpen
  \bibfield  {author} {\bibinfo {author} {\bibfnamefont {C.}~\bibnamefont
  {Andres}}, \bibinfo {author} {\bibfnamefont {L.}~\bibnamefont
  {Apolin\'ario}}, \bibinfo {author} {\bibfnamefont {N.}~\bibnamefont
  {Armesto}}, \bibinfo {author} {\bibfnamefont {A.}~\bibnamefont {Cordeiro}},
  \bibinfo {author} {\bibfnamefont {F.}~\bibnamefont {Dom\'inguez}}, \bibinfo
  {author} {\bibfnamefont {P.}~\bibnamefont {Guerrero-Rodr\'iguez}},\ and\
  \bibinfo {author} {\bibfnamefont {J.~G.}\ \bibnamefont {Milhano}},\
  }\bibfield  {title} {\bibinfo {title} {Gluon radiation from a {QCD} antenna
  with realistic parton-medium interactions},\ }\href@noop {} {\bibfield
  {journal} {\bibinfo  {journal} {arXiv:2607.00629 [hep-ph]}\ } (\bibinfo
  {year} {2026})}\BibitemShut {NoStop}%
\bibitem [{\citenamefont {Djordjevic}\ and\ \citenamefont
  {Djordjevic}(2012)}]{DjordjevicDjordjevic2012}%
  \BibitemOpen
  \bibfield  {author} {\bibinfo {author} {\bibfnamefont {M.}~\bibnamefont
  {Djordjevic}}\ and\ \bibinfo {author} {\bibfnamefont {M.}~\bibnamefont
  {Djordjevic}},\ }\bibfield  {title} {\bibinfo {title} {Generalization of
  radiative jet energy loss to non-zero magnetic mass},\ }\href
  {https://doi.org/10.1016/j.physletb.2012.02.019} {\bibfield  {journal}
  {\bibinfo  {journal} {Phys. Lett. B}\ }\textbf {\bibinfo {volume} {709}},\
  \bibinfo {pages} {229} (\bibinfo {year} {2012})},\ \Eprint
  {https://arxiv.org/abs/1105.4359} {1105.4359} \BibitemShut {NoStop}%
\end{thebibliography}%
\end{document}